\documentclass[preprint,12pt]{elsarticle}

\usepackage{amssymb}

\journal{***}

\begin{document}

\begin{frontmatter}



\title{A class of repeated-root constacyclic codes over $\mathbb{F}_{p^m}[u]/\langle u^e\rangle$ of
Type $2$}


\author{Yuan Cao$^{a, \ b}$, Yonglin Cao$^{a, \ \ast}$, Hai Q. Dinh$^{c,d,e}$, Fang-Wei Fu$^{f}$, Jian Gao$^{a}$, Songsak Sriboonchitta$^{g}$}

\address{$^{a}$School of Mathematics and  Statistics,
Shandong University of Technology, Zibo, Shandong 255091, China\\

\vskip 1mm $^{b}$ School of Mathematics and Statistics, Changsha University of Science and Technology,
Changsha, Hunan 410114, China\\
\vskip 1mm $^{c}$Division of Computational Mathematics and Engineering, Institute for Computational Science, Ton Duc Thang University, Ho Chi Minh City, Vietnam
\vskip 1mm $^{d}$Faculty of Mathematics and Statistics, Ton Duc Thang University, Ho Chi Minh City, Vietnam
\vskip 1mm $^{e}$Department of Mathematical Sciences, Kent State University, 4314 Mahoning Avenue, Warren, OH 44483, USA
\vskip 1mm $^{f}$Chern Institute of Mathematics and LPMC, Nankai University, Tianjin 300071, China
\vskip 1mm $^{g}$Faculty of Economics, Chiang Mai University, Chiang Mai 52000, Thailand}
\cortext[cor1]{corresponding author.  \\
E-mail addresses: yuancao@sdut.edu.cn (Yuan Cao), \ ylcao@sdut.edu.cn (Yonglin Cao),
\ hdinh@kent.edu (H. Q. Dinh), \ fwfu@nankai.edu.cn (F-W. Fu), dezhougaojian@163.com (J. Gao),
\ songsakecon@gmail.com (S. Sriboonchitta). }

\begin{abstract}
Let $\mathbb{F}_{p^m}$ be a finite field of cardinality $p^m$ where $p$ is an odd prime,
$n$ be a positive integer satisfying ${\rm gcd}(n,p)=1$,
and denote $R=\mathbb{F}_{p^m}[u]/\langle u^e\rangle$
where $e\geq 4$ be an even integer. Let $\delta,\alpha\in \mathbb{F}_{p^m}^{\times}$.
Then the class of $(\delta+\alpha u^2)$-constacyclic codes over $R$ is a
significant subclass of constacyclic codes over $R$ of Type 2.
For any integer $k\geq 1$, an explicit representation and a complete description for all distinct $(\delta+\alpha u^2)$-constacyclic codes over $R$ of length $np^k$ and their dual codes are given. Moreover,
formulas for the number of codewords in each code and the number of all such codes are provided
respectively. In particular,
all distinct $(\delta+\alpha u^2)$-contacyclic codes over $\mathbb{F}_{p^m}[u]/\langle u^{e}\rangle$ of length $p^k$ and their dual codes are presented precisely.
\end{abstract}


\end{frontmatter}


\section{Introduction}
\noindent
 Algebraic coding theory deals with the design of error-correcting and error-detecting codes for the reliable transmission
of information across noisy channel. The class of constacyclic codes plays a very significant role in
the theory of error-correcting codes.  It includes as a subclass of the important class of cyclic codes, which has been well studied since the late 1950's. Constacyclic codes also have practical applications as they can be efficiently encoded with simple shift registers. This family of codes is thus interesting for both theoretical and practical reasons.

\par
  Let $\Gamma$ be a commutative finite ring with identity $1\neq 0$, and $\Gamma^{\times}$ be the multiplicative group of invertible elements of
$\Gamma$. For any $a\in
\Gamma$, we denote by $\langle a\rangle_\Gamma$, or $\langle a\rangle$ for
simplicity, the ideal of $\Gamma$ generated by $a$, i.e. $\langle
a\rangle_\Gamma=a\Gamma=\{ab\mid b\in \Gamma\}$. For any ideal $I$ of $\Gamma$, we will identify the
element $a+I$ of the residue class ring $\Gamma/I$ with $a$ (mod $I$) for
any $a\in \Gamma$ in this paper.

\par
   A \textit{code} over $\Gamma$ of length $N$ is a nonempty subset ${\cal C}$ of $\Gamma^N=\{(a_0,a_1,\ldots$, $a_{N-1})\mid a_j\in\Gamma, \
j=0,1,\ldots,N-1\}$. The code ${\cal C}$
is said to be \textit{linear} if ${\cal C}$ is an $\Gamma$-submodule of $\Gamma^N$. All codes in this paper are assumed to be linear. The ambient space $\Gamma^N$ is equipped with the usual Euclidian inner product, i.e.
$[a,b]=\sum_{j=0}^{N-1}a_jb_j$, where $a=(a_0,a_1,\ldots,a_{N-1}), b=(b_0,b_1,\ldots,b_{N-1})\in \Gamma^N$,
and the \textit{dual code} is defined by ${\cal C}^{\bot}=\{a\in \Gamma^N\mid [a,b]=0, \forall b\in {\cal C}\}$.
If ${\cal C}^{\bot}={\cal C}$, ${\cal C}$ is called a \textit{self-dual code} over $\Gamma$.
   Let $\gamma\in \Gamma^{\times}$.
Then a linear code
${\cal C}$ over $\Gamma$ of length $N$ is
called a $\gamma$-\textit{constacyclic code}
if $(\gamma c_{N-1},c_0,c_1,\ldots,c_{N-2})\in {\cal C}$ for all
$(c_0,c_1,\ldots,c_{N-1})\in{\cal C}$. Particularly, ${\cal C}$ is
called a \textit{negacyclic code} if $\gamma=-1$, and ${\cal C}$ is
called a  \textit{cyclic code} if $\gamma=1$.

\par
  For any $a=(a_0,a_1,\ldots,a_{N-1})\in \Gamma^N$, let
$a(x)=a_0+a_1x+\ldots+a_{N-1}x^{N-1}\in \Gamma[x]/\langle x^N-\gamma\rangle$. We will identify $a$ with $a(x)$ in
this paper. Then ${\cal C}$ is a  $\gamma$-constacyclic code over $\Gamma$
of length $N$ if and only if ${\cal C}$ is an ideal of
the residue class ring $\Gamma[x]/\langle x^N-\gamma\rangle$, and the dual code ${\cal C}^{\bot}$ of ${\cal C}$ is a $\gamma^{-1}$-constacyclic code of length $N$ over
$\Gamma$, i.e. ${\cal C}^{\bot}$ is an ideal of the ring $\Gamma[x]/\langle
x^N-\gamma^{-1}\rangle$ (cf. [10] Propositions 2.4 and 2.5). The ring $\Gamma[x]/\langle x^N-\gamma\rangle$
is usually called the \textit{ambient ring} of $\gamma$-constacyclic codes over $\Gamma$
with length $N$.

\par
  Let $\mathbb{F}_{q}$ be a finite field of cardinality $q$, where
$q$ is power of a prime, and denote $R=\mathbb{F}_{q}[u]/\langle u^e\rangle
=\mathbb{F}_{q}+u\mathbb{F}_{q}+\ldots+u^{e-1}\mathbb{F}_{q}$ ($u^e=0$) where $e\geq 2$. Then
$R$ is a finite chain ring. As in Dinh et al [10], if
$$\gamma=\alpha_0+\alpha_ku^k+\ldots+\alpha_{e-1}u^{e-1}$$
where $\alpha_0,\alpha_k,\ldots,\alpha_{e-1}\in \mathbb{F}_{q}$ satisfying $\alpha_0\alpha_k\neq 0$, then $\gamma$
is called a unit in $R$ to be of \textit{Type $k$}. When $\gamma$ is a unit in $R$ of Type $k$, a $\gamma$-constacyclic code $\mathcal{C}$
of length $N$ over $R$ is said to be of \textit{Type $k$}. On the other hand,
$\mathcal{C}$ is called a \textit{simple-root constacyclic code} if
${\rm gcd}(q,N)=1$, and called a \textit{repeated-root constacyclic code} otherwise.

\par
  When $e=2$, there were a lot of literatures on linear codes, cyclic codes and
constacyclic codes of length $N$ over rings $\mathbb{F}_{p^m}[u]/\langle u^2\rangle
=\mathbb{F}_{p^m}+u\mathbb{F}_{p^m}$ for various prime $p$ and positive integers $m$ and $N$.
See [1--3], [11--17], [19], [22] and [24], for examples. In particular, an explicit representation for all $\alpha_0$-constacyclic codes
over $\mathbb{F}_{p^m}+u\mathbb{F}_{p^m}$ of arbitrary length and their dual codes are given in [3]
for any $\alpha_0\in \mathbb{F}_{p^m}^\times$, prime number $p$ and positive integer $m$.

\par
  When $e\geq 3$, the structures for repeated-root constacyclic codes of Type $1$ over $R$ had been studied by many literatures. For examples,
Kai et al. [20] investigated $(1+\lambda u)$-constacyclic codes of arbitrary length over $\mathbb{F}_p[u]/\langle u^k\rangle$, where $\lambda\in \mathbb{F}_p^\times$. Cao [4] generalized
these results to $(1+w\gamma)$-constacyclic codes of arbitrary length over an arbitrary finite
chain ring $\Gamma$, where $w$ is a unit of $\Gamma$ and $\gamma$ generates the unique maximal ideal of $\Gamma$
with nilpotency index $e\geq 2$. Hence every constacyclic code of Type 1 over any finite chain ring is
a one-generator ideal of the ambient ring.

\par
  When $e\geq 3$ and $k\geq 2$, there were 	
fewer literatures on repeated-root constacyclic codes over $R$ of Type $k$.

\par
  For repeated-root constacyclic codes over $R$ of Type $2$,
in the case of $e=3$, Sobhani [23] determined the structure of $(\delta+\alpha u^2)$-constacyclic codes
of length $p^k$ over $\mathbb{F}_{p^m}[u]/\langle u^3\rangle$, where $\delta,\alpha\in \mathbb{F}_{p^m}^{\times}$.
  When $e=4$ and ${\rm gcd}(q,n)=1$, in [5] for any $\delta,\alpha\in \mathbb{F}_{q}^{\times}$,
an explicit representation for all distinct $(\delta+\alpha u^2)$-constacyclic codes over the ring $\mathbb{F}_{q}[u]/\langle u^4\rangle$ of
length $n$ is given, and the dual code for each of these codes is determined. For the case of $q=2^m$ and $\delta=1$, all self-dual $(1+\alpha u^2)$-constacyclic codes over $R$ of
length $n$ are provided.

\par
   Let $e=4$. When $p=3$, in [6] an explicit representation for all distinct $(\delta+\alpha u^2)$-constacyclic codes over $\mathbb{F}_{3^m}[u]/\langle u^4\rangle$ of length
$3n$ was given, where ${\rm gcd}(3,n)=1$. Formulas for the number of all such codes and the number of codewords in
each code are provided respectively, and the dual code for each of these codes
was determined explicitly. When $p=2$, in [7] a representation and
enumeration formulas for all distinct $(\delta+\alpha u^2)$-constacyclic codes  over $\mathbb{F}_{2^m}[u]/\langle u^4\rangle$ of length $2n$ were presented explicitly, where $n$ is odd.

\par
  Motivated by those, we generalize the approach used in [6] to determine the structures
of repeated-root $(\delta+\alpha u^2)$-constacyclic codes over $\mathbb{F}_{p^m}[u]/\langle u^e\rangle$
for any $\delta,\alpha\in \mathbb{F}_{p^m}^\times$.
This class is a significant subclass of constacyclic codes over finite chain rings of Type 2.
We give a precise representation
and a complete classification for this class of constacyclic codes and their dual codes in this paper.
  From now on, we adopt the following notations.

\vskip 3mm \noindent
   {\bf Notation 1.1} Let $p$ be an odd prime number, $m, n, k$ be positive integers
 satisfying ${\rm gcd}(n,p)=1$. For any even integer $e\geq 4$ and nonzero elements $\delta,\alpha\in \mathbb{F}_{p^m}$, we denote

\vskip 2mm \par
   $\bullet$ $e=2\lambda$ where $\lambda\geq 2$ being an integer.

\vskip 2mm \par
   $\bullet$ $R=\mathbb{F}_{p^m}[u]/\langle u^e\rangle=\mathbb{F}_{p^m}
+u\mathbb{F}_{p^m}+u^2\mathbb{F}_{p^m}+\ldots+u^{e-1}\mathbb{F}_{p^m}$ ($u^e=0$), which is a finite chain ring
with the unique maximal ideal $uR$.

\vskip 2mm \par
   $\bullet$ $\mathcal{A}=\mathbb{F}_{p^m}[x]/\langle(x^{np^k}-\delta)^\lambda\rangle$, which is a finite principal ideal ring, $|\mathcal{A}|=p^{\lambda mnp^k}$, and
$\mathcal{A}=\{\sum_{i=0}^{\lambda np^k-1}a_ix^i\mid a_i\in \mathbb{F}_{p^m}, \ i=0,1,\ldots,\lambda np^k-1\}$
in which the arithmetics are done modulo $(x^{np^k}-\delta)^\lambda$.

\vskip 2mm \par
   $\bullet$ $\mathcal{A}[u]/\langle u^2-\alpha^{-1}(x^{np^k}-\delta)\rangle=\mathcal{A}+u\mathcal{A}$ ($u^2=\alpha^{-1}(x^{np^k}-\delta)$), where
$\mathcal{A}+u\mathcal{A}=\{\xi_0+u\xi_1\mid \xi_0,\xi_1\in \mathcal{A}\}$, $|\mathcal{A}+u\mathcal{A}|=p^{2\lambda mnp^k}$
 and the operations are defined by

\vskip 2mm \par
  $\diamond$ $(\xi_0+u\xi_1)+(\eta_0+u\eta_1)=(\xi_0+\eta_0)+u(\xi_1+\eta_1)$,

\vskip 2mm \par
  $\diamond$ $(\xi_0+u\xi_1)(\eta_0+u\eta_1)=\left(\xi_0\eta_0+\alpha^{-1}(x^{np^k}-\delta)\xi_1\eta_1\right)+u(\xi_0\eta_1+\xi_1\eta_0)$,

\vskip 2mm \noindent
  for all $\xi_0,\xi_1,\eta_0,\eta_1\in \mathcal{A}$.

\vskip 2mm \par
   $\bullet$ $\delta_0\in \mathbb{F}_{p^m}^{\times}$ satisfying $\delta_0^{p^k}=\delta$. (Since $\delta\in \mathbb{F}_{p^m}^{\times}$ and $|\mathbb{F}_{p^m}^{\times}|=p^m-1$, there is a unique
$\delta_0\in \mathbb{F}_{p^m}^{\times}$ such that $\delta_0^{p^k}=\delta$).

\vskip 2mm\par
    The present paper is organized as follows. In Section 2, we construct a ring isomorphism from $\mathcal{A}+u\mathcal{A}$ onto
$R[x]/\langle x^{np^k}-(\delta+\alpha u^2)\rangle$ first. Then by the Chinese remainder theorem, we give a direct sum decomposition for $\mathcal{A}+u\mathcal{A}$, which induces a direct sum decomposition for any $(\delta+\alpha u^2)$-constacyclic code over $R$ of length $np^k$.
In Section 3, we determine the direct summands and provide an explicit representation for each $(\delta+\alpha u^2)$-contacyclic code over $R$ of length $np^k$. Using this representation, we give formulas to count the number of codewords in each code and the number of all such codes respectively.
Then we give the dual code of each $(\delta+\alpha u^2)$-contacyclic code of length $np^k$ over $R$ in Section 4.
In Section 5. we determine all distinct $(\delta+\alpha u^2)$-contacyclic codes over $R$ of length $np^k$ when $x^n-\delta_0$ is irreducible in $\mathbb{F}_{p^m}[x]$ and $\delta=\delta_0^{p^k}$. In particular,
we list all distinct $(\delta+\alpha u^2)$-contacyclic codes and their dual codes over $R$ of length $p^k$ explicitly.


\section{Direct sum decomposition of $(\delta+\alpha u^2)$-constacyclic codes over $R$ of length $np^k$}
\noindent
  In this section, we will construct a specific isomorphism of rings from $\mathcal{A}+u\mathcal{A}$ onto
$R[x]/\langle x^{np^k}-(\delta+\alpha u^2)\rangle$. Hence we obtain a one-to-one correspondence
between the set of ideals in the ring $\mathcal{A}+u\mathcal{A}$ onto the set of ideals in the ring
$R[x]/\langle x^{np^k}-(\delta+\alpha u^2)\rangle=\{\sum_{i=0}^{np^k-1}r_ix^i\mid
r_0,r_1,\ldots,r_{np^k-1}\in R\}$
in which the arithmetics are done modulo $x^{np^k}-(\delta+\alpha u^2)$.
Then we provide a direct sum decomposition
for any $(\delta+\alpha u^2)$-constacyclic code over $R$ of length $np^k$.

\par
  Let $\xi_0+u\xi_1\in \mathcal{A}+u\mathcal{A}$ where $\xi_0,\xi_1\in \mathcal{A}$. It is clear that
$\xi_0$ can be uniquely expressed
as $\xi_0=\xi_0(x)$ where $\xi_0(x)\in \mathbb{F}_{p^m}[x]$ satisfying ${\rm deg}(\xi_0(x))<\lambda np^k$ (we will write ${\rm deg}(0)=-\infty$ for convenience). Dividing $\xi_0(x)$ by $\alpha^{-1}(x^{np^k}-\delta)$ iteratively, we obtain
a unique ordered $\lambda$-tuple $(a_0(x),a_2(x),\ldots,a_{2(\lambda-1)}(x))$ of polynomials in $\mathbb{F}_{p^m}[x]$ such that
$$\xi_0=\xi_0(x)=\sum_{j=0}^{\lambda-1}\left(\alpha^{-1}(x^{np^k}-\delta)\right)^{j}a_{2j}(x)$$
and ${\rm deg}(a_{2j}(x))<np^k$ for all $j=0,1,\ldots,\lambda-1$.
Similarly, there is a unique ordered $\lambda$-tuple $(a_1(x),a_3(x),\ldots,a_{2\lambda-1}(x))$ of polynomials in $\mathbb{F}_{p^m}[x]$ such that
$$\xi_1=\xi_1(x)=\sum_{j=0}^{\lambda-1}\left(\alpha^{-1}(x^{np^k}-\delta)\right)^{j}a_{2j+1}(x)$$
and ${\rm deg}(a_{2j+1}(x))<np^k$ for all $j=0,1,\ldots,\lambda-1$.
Assume that
$$a_k(x)=\sum_{i=0}^{np^k-1}a_{i,k}x^i \ {\rm where} \ a_{i,k}\in \mathbb{F}_{p^m},
\ 0\leq i\leq np^k-1 \ {\rm and} \ 0\leq k\leq 2\lambda-1.$$
Then $\xi_0+u\xi_1$ can be written as a product of matrices:
$$\xi_0+u\xi_1=(1,x,\ldots,x^{np^k-1})M\left(\begin{array}{c}1 \cr u \cr \alpha^{-1}(x^{np^k}-\delta)\cr u\alpha^{-1}(x^{np^k}-\delta)\cr \ldots\cr (\alpha^{-1}(x^{np^k}-\delta))^{\lambda-1}\cr u(\alpha^{-1}(x^{np^k}-\delta))^{\lambda-1}\end{array}\right),$$
where $M=\left(a_{i,k}\right)_{0\leq i\leq np^k-1, 0\leq k\leq 2\lambda-1}$ is an $np^k\times 2\lambda$ matrix over $\mathbb{F}_{p^m}$. Define
\begin{eqnarray*}
\Psi(\xi_0+u\xi_1)&=&\left(1,x,\ldots,x^{np^k-1}\right)M\left(\begin{array}{c}1 \cr u \cr \ldots\cr u^{2\lambda-2}\cr u^{2\lambda-1}\end{array}\right)
  =\sum_{i=0}^{np^k-1}r_ix^i,
\end{eqnarray*}
where $r_i=\sum_{k=0}^{2\lambda-1}u^ka_{i,k}\in R$ for all $i=0,1,\ldots,np^k-1$. Then it is clear that $\Psi$ is a bijection from
$\mathcal{A}+u\mathcal{A}$ onto $R[x]/\langle x^{np^k}-(\delta+\alpha u^2)\rangle$. Furthermore, from
$u^2=\alpha^{-1}(x^{np^k}-\delta)$, $(x^{np^k}-\delta)^\lambda=0$ in $\mathcal{A}+u\mathcal{A}$ and $x^{np^k}-(\delta+\alpha u^2)=0$
in $R[x]/\langle x^{np^k}-(\delta+\alpha u^2)\rangle$ we deduce the following conclusion.

\vskip 3mm \noindent
   {\bf Theorem 2.1} \textit{Using the notations above, $\Psi$ is a ring isomorphism from
$\mathcal{A}+u\mathcal{A}$ onto $R[x]/\langle x^{np^k}-(\delta+\alpha u^2)\rangle$}.

\vskip 3mm\noindent
  {\bf Proof.} Both $\mathcal{A}+u\mathcal{A}$ and $R[x]/\langle x^{np^k}-(\delta+\alpha u^2)\rangle$ are $\mathbb{F}_{p^m}$-algebras of dimension $2\lambda np^k$. In fact, $\{1,x,\ldots,x^{\lambda np^k-1},u,ux,\ldots,ux^{\lambda np^k-1}\}$ is an $\mathbb{F}_{p^m}$-basis of $\mathcal{A}+u\mathcal{A}$,
and $\cup_{i=0}^{2\lambda-1}\{u^i,u^ix,\ldots,u^ix^{np^k-1}\}$
  is an $\mathbb{F}_{p^m}$-basis of
$R[x]/\langle x^{np^k}-(\delta+\alpha u^2)\rangle$. It is clear that
$\Psi$ is an $\mathbb{F}_{p^m}$-linear space isomorphism from $\mathcal{A}+u\mathcal{A}$ onto $R[x]/\langle x^{np^k}-(\delta+\alpha u^2)\rangle$, and $\Psi$ is completely determined by:
\begin{equation}
\Psi(x^i)=x^i \ {\rm if} \ 0\leq i\leq np^k-1, \ \Psi\left(\alpha^{-1}(x^{np^k}-\delta)\right)=u^2 \ {\rm and} \ \Psi(u)=u.
\end{equation}
These imply that

\par
  $\diamondsuit$ $\Psi(b(x))=b(x)$ for all $b(x)\in \mathbb{F}_{p^m}[x]$ satisfying
${\rm deg}(b(x))\leq np^k-1$;

\par
  $\diamondsuit$ $\Psi(x^{np^k})=\alpha\Psi(\alpha^{-1}(x^{np^k}-\delta))+\delta
 =\delta+\alpha u^2$.

\par
  Let $\eta_0+u\eta_1\in \mathcal{A}+u\mathcal{A}$ where
$\eta_s=\sum_{j=0}^{\lambda-1}(\alpha^{-1}(x^{np^k}-\delta))^jb_{2j+s}(x)\in \mathcal{A}$
with $b_{2j+s}(x)\in\mathbb{F}_{p^m}[x]$ having degree less than $np^k$ for $s=0,1$ and $j=0,1,\ldots,\lambda-1$.
Then it is clear that $\Psi((\xi_0+u\xi_1)+(\eta_0+u\eta_1))=\Psi(\xi_0+u\xi_1)+\Psi(\eta_0+u\eta_1)$.
Moreover, by $(x^{np^k}-\delta)^\lambda=0$ in $\mathcal{A}$ and $u^{2\lambda}=0$ in $R$, we have
\begin{eqnarray*}
&&\Psi((\xi_0+u\xi_1)(\eta_0+u\eta_1))\\
&=&\Psi\left(\left(\xi_0\eta_0+\alpha^{-1}(x^{np^k}-\delta)\xi_1\eta_1\right)+u(\xi_0\eta_1+\xi_1\eta_0)\right)\\
&=&\Psi(\sum_{i+j<\lambda, \ 0\leq i,j\leq\lambda-1}(\alpha^{-1}(x^{np^k}-\delta))^{i+j}\cdot(a_{2i}(x)b_{2j}(x) \\
&&+\alpha^{-1}(x^{np^k}-\delta)a_{2i+1}(x)b_{2j+1}(x)
 +u(a_{2i}(x)b_{2j+1}(x)+a_{2i+1}(x)b_{2j}(x))))  \\
&=&\sum_{k=0}^{2\lambda-1}u^k(\sum_{s+t=k}a_s(x)b_t(x)) \ ({\rm mod} \ \alpha^{-1}(x^{np^k}-\delta)-u^2) \\
&=&(\sum_{s=0}^{2\lambda-1}u^sa_s(x))(\sum_{t=0}^{2\lambda-1}u^ta_t(x)) \ ({\rm mod} \ x^{np^k}-(\delta+\alpha u^2))\\
&=&\Psi(\xi_0+u\xi_1)\cdot\Psi(\eta_0+u\eta_1).
\end{eqnarray*}
Hence $\Psi$ is a ring isomorphism from
$\mathcal{A}+u\mathcal{A}$ onto $R[x]/\langle x^{2n}-(\delta+\alpha u^2)\rangle$.
\hfill $\Box$

\vskip 3mm\par
   By Theorem 2.1, $\Psi$ induces
a one-to-one correspondence
between the set of ideals in the ring $\mathcal{A}+u\mathcal{A}$ onto the set of ideas in the ring
$R[x]/\langle x^{np^k}-(\delta+\alpha u^2)\rangle$. Therefore,
in order to determine all distinct $(\delta+\alpha u^2)$-constacyclic codes over $R$ of length $np^k$,
it is sufficient to list all distinct
ideals of $\mathcal{A}+u\mathcal{A}$.

\par
   Now, we investigate structures and properties
of the rings $\mathcal{A}$ and $\mathcal{A}+u\mathcal{A}$.
   As $\delta\in \mathbb{F}_{p^m}^{\times}$ satisfying $\delta_0^{p^k}=\delta$, we have
$x^{np^k}-\delta=(x^n-\delta_0)^{p^k}$ in $\mathbb{F}_{p^m}[x]$.
By ${\rm gcd}(n,p)=1$,
    there are pairwise coprime monic
irreducible polynomials $f_1(x),\ldots, f_r(x)$ in $\mathbb{F}_{p^m}[x]$ such that $x^{n}-\delta_0=f_1(x)\ldots f_r(x)$. This
implies
\begin{equation}
(x^{np^k}-\delta)^\lambda=(x^n-\delta_0)^{\lambda p^k}=f_1(x)^{\lambda p^k}\ldots f_r(x)^{\lambda p^k}.
\end{equation}
For any integer $j$, $1\leq j\leq r$, we assume ${\rm deg}(f_j(x))=d_j$ and denote $F_j(x)=\frac{x^{n}-\delta_0}{f_j(x)}$.
Then $F_j(x)^{\lambda p^k}=\frac{(x^{np^k}-\delta)^\lambda}{f_j(x)^{\lambda p^k}}$ and ${\rm gcd}(F_j(x)^{\lambda p^k},f_j(x)^{\lambda p^k})=1$. So there exist $g_j(x),h_j(x)\in \mathbb{F}_{p^m}[x]$ such that
\begin{equation}
g_j(x)F_j(x)^{\lambda p^k}+h_j(x)f_j(x)^{\lambda p^k}=1.
\end{equation}

\par
  In the rest of this paper, we adopt the following notations.

\vskip 3mm \noindent
  {\bf Notation 2.2} Let $j$ be an integer satisfying  $1\leq j\leq r$.

\noindent
  $\bullet$ Let $\varepsilon_j(x)\in \mathcal{A}$ be defined by
$$\varepsilon_j(x)\equiv g_j(x)F_j(x)^{\lambda p^k}=1-h_j(x)f_j(x)^{\lambda p^k} \ ({\rm mod} \ (x^{np^k}-\delta)^\lambda).$$
\noindent
  $\bullet$ Denote
$\mathcal{K}_j=\mathbb{F}_{p^m}[x]/\langle f_j(x)^{\lambda p^k}\rangle=\{\sum_{i=0}^{d_j\lambda p^k-1}a_ix^i\mid
a_i\in \mathbb{F}_{p^m}, 0\leq i<d_j\lambda p^k\}$ in which the arithmetics are done modulo $f_j(x)^{\lambda p^k}$.

\vskip 3mm \par
  Then from Chinese remainder theorem for commutative rings, we deduce the following lemma about the structure and
properties of the ring $\mathcal{A}$.

\vskip 3mm
\noindent
  {\bf Lemma 2.3} \textit{Using the notations above, we have the following  decomposition
via idempotents}:

\vskip 2mm \par
  (i) \textit{$\varepsilon_1(x)+\ldots+\varepsilon_r(x)=1$, $\varepsilon_j(x)^2=\varepsilon_j(x)$
and $\varepsilon_j(x)\varepsilon_l(x)=0$  in the ring $\mathcal{A}$ for all $1\leq j\neq l\leq r$}.

\vskip 2mm \par
  (ii) \textit{For any $a_j(x)\in \mathcal{K}_j$ with $j=1,\ldots,r$, define
$$\varphi(a_1(x),\ldots,a_r(x))=\sum_{j=1}^r\varepsilon_j(x)a_j(x) \ ({\rm mod} \ (x^{np^k}-\delta)^\lambda).$$
Then
$\varphi$ is a ring isomorphism from $\mathcal{K}_1\times\ldots\times\mathcal{K}_r$ onto $\mathcal{A}$}.

\vskip 3mm \par
    In order to study the structure of the ring $\mathcal{A}+u\mathcal{A}$ ($u^2=\alpha^{-1}(x^{np^k}-\delta)$),
we need the following lemma.

\vskip 3mm
\noindent
  {\bf Lemma 2.4} \textit{Let $1\leq j\leq r$ and denote
$\omega_j=\alpha^{-1}F_j(x)^{p^k} \ ({\rm mod} \ f_j(x)^{\lambda p^k}).$
Then $\omega_j$ is an invertible element of $\mathcal{K}_j$ and satisfies
$$\alpha^{-1}(x^{np^k}-\delta)=\omega_jf_j(x)^{p^k} \ ({\rm mod} \ \ f_j(x)^{\lambda p^k}).$$
Hence $\alpha^{-1}(x^{np^k}-\delta)=\omega_jf_j(x)^{p^k}$ in the ring
$\mathcal{K}_j$}.

\vskip 3mm
\noindent
  {\bf Proof.} Since $\omega_j\in \mathcal{K}_j$ satisfying $\omega_j\equiv\alpha^{-1}F_j(x)^{p^k}$ (mod $f_j(x)^{\lambda p^k}$),
by Equation (3) it follows that
\begin{eqnarray*}
\left(\alpha g_j(x)F_j(x)^{(\lambda-1)p^k}\right)\omega_j
&\equiv&\left(\alpha g_j(x)F_j(x)^{(\lambda-1)p^k}\right)\left(\alpha^{-1}F_j(x)^{p^k}\right)\\
 &=&g_j(x)F_j(x)^{\lambda p^k}=1-h_j(x)f_j(x)^{\lambda p^k}\\
 &\equiv& 1 \ ({\rm mod} \ f_j(x)^{\lambda p^k}).
\end{eqnarray*}
This implies $(\alpha g_j(x)F_j(x)^{(\lambda-1)p^k})\omega_j=1$ in $\mathcal{K}_j$. Hence $\omega_j\in \mathcal{K}_j^{\times}$
and $\omega_j^{-1}=\alpha g_j(x)F_j(x)^{(\lambda-1)p^k}$ (mod $f_j(x)^{\lambda p^k}$).
Then from $x^{np^k}-\delta=f_1(x)^{ p^k}\ldots f_r(x)^{ p^k}$ and $F_j(x)^{p^k}=\frac{(x^{n}-\delta_0)^{p^k}}{f_j(x)^{p^k}}=\frac{x^{np^k}-\delta}{f_j(x)^{p^k}}$, we deduce
the equality
$\alpha^{-1}(x^{np^k}-\delta)=\alpha^{-1}F_j(x)^{p^k}f_j(x)^{p^k}=
\omega_jf_j(x)^{p^k}$ in $\mathcal{K}_j$.
\hfill $\Box$

\vskip 3mm
\par
   Now, we determine the structure of $\mathcal{A}+u\mathcal{A}$ by the following lemma.

\vskip 3mm
\noindent
  {\bf Lemma 2.5} \textit{Let $1\leq j\leq r$. Using the notations in Lemma 2.4, we denote}
$$\mathcal{K}_j[u]/\langle u^2-\omega_jf_j(x)^{p^k}\rangle=\mathcal{K}_j+u\mathcal{K}_j \ (u^2=\omega_jf_j(x)^{p^k}).$$
\textit{For any $\beta_j,\gamma_j\in \mathcal{K}_j$, $j=1,\ldots,r$, define}
\begin{eqnarray*}
 &&\Phi(\beta_1+u\gamma_1,\ldots,\beta_r+u\gamma_r)
 =\varphi(\beta_1,\ldots,\beta_r)+u\varphi(\gamma_1,\ldots,\gamma_r)\\
&=&\sum_{j=1}^r\varepsilon_j(x)\left(\beta_j+u\gamma_j\right) \ ({\rm mod} \ (x^{np^k}-\delta)).
\end{eqnarray*}
\textit{Then
$\Phi$ is a ring isomorphism from $(\mathcal{K}_1+u\mathcal{K}_1)\times\ldots\times(\mathcal{K}_r+u\mathcal{K}_r)$
onto $\mathcal{A}+u\mathcal{A}$}.

\vskip 3mm\noindent
  {\bf Proof.} The ring isomorphism $\varphi:\mathcal{K}_1\times\ldots\times\mathcal{K}_r\rightarrow\mathcal{A}$
defined in Lemma 2.3(ii) can be extended to a polynomial ring isomorphism $\Phi_0$ from $(\mathcal{K}_1\times\ldots\times\mathcal{K}_r)[u]=\mathcal{K}_1[u]\times\ldots\times\mathcal{K}_r[u]$ onto $\mathcal{A}[u]$ in
the natural way that
\begin{eqnarray*}
&&\Phi_0(\sum_t\beta_{1,t}u^t,\ldots,\sum_t\beta_{r,t}u^t)\\
&=&\sum_t\left(\sum_{j=1}^r\varepsilon_j(x)\beta_{j,t}\right)u^t
=\sum_t\varphi\left(\beta_{1,t},\ldots,\beta_{r,t}\right)u^t
\end{eqnarray*}
for all $\beta_{j,t}\in \mathcal{K}_j$. From this, by Lemma 2.3 (ii) and Lemma 2.4  we deduce
\begin{eqnarray*}
&&\Phi_0\left(u^2-\omega_1f_1(x)^{p^k}, \ldots, u^2-\omega_rf_r(x)^{p^k}\right)\\
  &=&(\sum_{j=1}^r\varepsilon_j(x))u^2-\sum_{j=1}^r\varepsilon_j(x)\omega_jf_j(x)^{p^k}
  =u^2-\alpha^{-1}(x^{np^k}-\delta).
\end{eqnarray*}
Therefore, by classical ring theory we conclude that $\Phi_0$ induces a surjective ring homomorphism $\Phi$ from $\frac{\mathcal{K}_1[u]}{\langle u^2-\omega_1f_1(x)^{p^k}\rangle}\times\ldots\times
\frac{\mathcal{K}_r[u]}{\langle u^2-\omega_rf_r(x)^{p^k}\rangle}$
onto $\mathcal{A}[u]/\langle u^2-\alpha^{-1}(x^{np^k}-\delta)\rangle$ that is defined as in the lemma. From this and by
\begin{eqnarray*}
&&\left|\frac{\mathcal{K}_1[u]}{\langle u^2-\omega_1f_1(x)^{p^k}\rangle}\times\ldots\times
\frac{\mathcal{K}_r[u]}{\langle u^2-\omega_rf_r(x)^{p^k}\rangle}\right|\\
 &=&\prod_{j=1}^r|\mathcal{K}_j[u]/\langle u^2-\omega_jf_j(x)^{p^k}\rangle|=\prod_{j=1}^r|\mathcal{K}_j|^{2}=\prod_{j=1}^r(p^{md_j \lambda p^k})^2\\
 &=&p^{2m \lambda p^k\sum_{j=1}^rd_j}=p^{2m \lambda p^kn}=(p^{mn\lambda p^k})^2=|\mathcal{A}|^2\\
 &=&|\mathcal{A}[u]/\langle u^2-\alpha^{-1}(x^{np^k}-\delta)\rangle|,
\end{eqnarray*}
we deduce that $\Phi$ is a ring isomorphism. Finally, the conclusion follows from
$\mathcal{A}[u]/\langle u^2-\alpha^{-1}(x^{np^k}-\delta)\rangle=\mathcal{A}+u\mathcal{A}$ by Notation 1.1 and
$\mathcal{K}_j[u]/\langle u^2-\omega_jf_j(x)^{p^k}\rangle=\mathcal{K}_j+u\mathcal{K}_j$ for all $j=1,\ldots,r$.
\hfill $\Box$

\vskip 3mm \noindent
   {\bf Lemma 2.6} \textit{For any integer $j$, $1\leq j\leq r$, denote $e_j(x)=\Psi(\varepsilon_j(x))\in R[x]/\langle x^{np^k}-(\delta+\alpha u^2)\rangle$. Then}

\vskip 2mm\par
   (i)  \textit{$e_1(x)+\ldots+e_r(x)=1$, $e_j(x)^2=e_j(x)$
and $e_j(x)e_l(x)=0$  in the ring $R[x]/\langle x^{np^k}-(\delta+\alpha u^2)\rangle$ for all $1\leq j\neq l\leq r$}.

\vskip 2mm\par
   (ii) \textit{Write $\varepsilon_j(x)=\sum_{s=0}^{\lambda-1}\left(\alpha^{-1}(x^{np^k}-\delta)\right)^se_{j,s}(x)$ where
$e_{j,s}(x)\in \mathbb{F}_{p^m}[x]$ satisfying ${\rm deg}(e_{j,s}(x))\leq np^k-1$ for $s=0,1,\ldots,\lambda-1$.
Then}
$$e_j(x)=e_{j,0}(x)+u^2e_{j,1}(x)+\ldots+u^{2(\lambda-1)}e_{j,\lambda-1}(x).$$

\vskip 3mm \noindent
  {\bf Proof.} (i) By Theorem 2.1, $\Psi$ is a ring isomorphism from
$\mathcal{A}+u\mathcal{A}$ onto $R[x]/\langle x^{np^k}-(\delta+\alpha u^2)\rangle$. Then the conclusions
follow from Lemma 2.3(i).

\par
  (ii) As $\mathcal{A}=\mathbb{F}_{p^m}[x]/\langle (x^{np^k}-\delta)^\lambda\rangle
=\mathbb{F}_{p^m}[x]/\langle (\alpha^{-1}(x^{np^k}-\delta))^\lambda\rangle$ and $\varepsilon_j(x)\in \mathcal{A}$,
there is a unique ordered $\lambda$-tuple $(e_{j,0}(x),e_{j,1}(x),\ldots,e_{j,\lambda-1}(x))$ of polynomials in $\mathbb{F}_{p^m}[x]$,
where ${\rm deg}(e_{j,s}(x))\leq np^k-1$ for $s=0,1,\ldots,\lambda-1$, such that $\varepsilon_j(x)=\sum_{s=0}^{\lambda-1}\left(\alpha^{-1}(x^{np^k}-\delta)\right)^se_{j,s}(x)$. Then by Theorem 2.1 and Equation (1),
we have $e_j(x)=\Psi(\varepsilon_j(x))
=\sum_{s=0}^{\lambda-1}\left(\Psi(\alpha^{-1}(x^{np^k}-\delta))\right)^s\Psi(e_{j,s}(x))$. This implies  $e_j(x)=\sum_{s=0}^{\lambda-1}u^{2s}e_{j,s}(x)$ by Equation (1).
\hfill $\Box$

\vskip 3mm \par
     Finally, we give a direct sum decomposition for any
$(\delta+\alpha u^2)$-constacyclic code over $R$ of length $np^k$.

\vskip 3mm \noindent
   {\bf Theorem 2.7} \textit{Let $\mathcal{C}$ be a $(\delta+\alpha u^2)$-constacyclic code of length $np^k$ over $R$.
Then for each integer $j$, $1\leq j\leq r$, there is a unique ideal $C_j$ of the ring $\mathcal{K}_j+u\mathcal{K}_j$ $(u^2=\omega_jf_j(x)^{p^k})$ such that}
$$\mathcal{C}=\bigoplus_{j=1}^r\Psi(\varepsilon_j(x)C_j)=\sum_{j=1}^re_j(x)\Psi(C_j) \
({\rm mod} \ x^{np^k}-(\delta+\alpha u^2)),$$
\textit{where $C_j$ is regarded as a subset of $\mathcal{A}+u\mathcal{A}$ and
$\Psi(C_j)=\{\Psi(\xi)\mid \xi\in C_j\}$ for all $j=1,\ldots,r$}.

\par
  \textit{Moreover, the number of codewords in $\mathcal{C}$ is equal to $|\mathcal{C}|=\prod_{j=1}^r|C_j|$}.

\vskip 3mm \noindent
   {\bf Proof.} By Theorem 2.1 and Lemma 2.5, we see that $\Psi\circ \Phi$ is a ring isomorphism from
$(\mathcal{K}_1+u\mathcal{K}_1)\times\ldots\times(\mathcal{K}_r+u\mathcal{K}_r)$ onto $R[x]/\langle x^{np^k}-(\delta+\alpha u^2)\rangle$.
Since $\mathcal{C}$ is an ideal of $R[x]/\langle x^{np^k}-(\delta+\alpha u^2)\rangle$, there is a unique ideal $C_j$ of
$\mathcal{K}_j+u\mathcal{K}_j$ for each $j=1,\ldots,r$, such that
\begin{eqnarray*}
\mathcal{C}&=&(\Psi\circ \Phi)(C_1\times\ldots\times C_r)=\Psi\left(\Phi\{(\xi_1,\ldots,\xi_r)\mid \xi_j\in C_j, \ j=1,\ldots,r\}\right)\\
     &=&\Psi\left(\left\{\sum_{j=1}^r\varepsilon_j(x)\xi_j\mid \xi_j\in C_j, \ j=1,\ldots,r\right\}\right)
     =\Psi\left(\bigoplus_{j=1}^r\varepsilon_j(x)C_j\right).
\end{eqnarray*}
By Lemma 2.5, $\varepsilon_j(x)C_j=\{\varepsilon_j(x)\xi\mid \xi\in C_j\}\subseteq\mathcal{A}+u\mathcal{A}$.
As $C_j\subseteq\mathcal{K}_j+u\mathcal{K}_j$, each element $\xi\in C_j$ can be uniquely repressed as
$\xi=a(x)+ub(x)$ where $a(x),b(x)\in\mathbb{F}_{p^m}[x]$ satisfying ${\rm deg}(a(x)), {\rm deg}(b(x))<d_j\lambda p^k\leq n\lambda p^k$.
Here we regard $a(x)+ub(x)$ as an element in $\mathcal{A}+u\mathcal{A}$, and obtain
$\Psi(\varepsilon_j(x)\xi)=\Psi(\varepsilon_j(x))\Psi(a(x)+ub(x))=e_j(x)\Psi(\xi)$ by Theorem 2.1. Hence $\mathcal{C}=\bigoplus_{j=1}^re_j(x)\Psi(C_j)$.
\hfill $\Box$

\vskip 3mm\par
   By Theorem 2.7, in order to determine all $(\delta+\alpha u^2)$-constacyclic code of length $np^k$ over $R$
it is sufficient to list all distinct ideals of the ring $\mathcal{K}_j+u\mathcal{K}_j$ for all $j=1,\ldots,r$.



\section{Representation for all distinct ideals of the ring $\mathcal{K}_j+u\mathcal{K}_j$}
\noindent
In this section, we use module theory over finite chain rings to give a precise representation for all distinct ideals of the ring $\mathcal{K}_j+u\mathcal{K}_j$ and enumerate them, where $\mathcal{K}_j=\mathbb{F}_{p^m}[x]/\langle f_j(x)^{\lambda p^k}\rangle$ and ${\rm deg}(f_j(x))=d_j$,
 for all $j=1,\ldots,r$.

\par
   Let $1\leq j\leq r$. In order to simplify the notations and expressions, we adopt the following notations in the rest of this paper:

\vskip 2mm
\noindent
  $\bullet$ $\pi_j=f_j(x)$;

\vskip 2mm
\noindent
  $\bullet$ ${\cal T}_j=\{\sum_{i=0}^{d_j-1}t_ix^i\mid t_0,t_1,\ldots,t_{d_j-1}\in \mathbb{F}_{p^m}\}\subset \mathcal{K}_j$;

\vskip 2mm
\noindent
  $\bullet$ ${\cal F}_j=\mathbb{F}_{p^m}[x]/\langle f_j(x)\rangle=\{\sum_{i=0}^{d_j-1}t_ix^i\mid t_0,t_1,\ldots,t_{d_j-1}\in \mathbb{F}_{p^m}\}$ in which the arithmetics are done modulo $f_j(x)$.

\par
  Since $f_j(x)$ is irreducible in $\mathbb{F}_{p^m}[x]$, ${\cal F}_j$ is an extension field
of $\mathbb{F}_{p^m}$ with $p^{md_j}$ elements. As $\lambda p^k>1$, ${\cal F}_j$ is not a subfield
of the ring $\mathcal{K}_j$ in which the arithmetics are done modulo $f_j(x)^{\lambda p^k}$. In this paper,
we will identify ${\cal F}_j$ with ${\cal T}_j$ as sets, i.e. we regard ${\cal F}_j$ as a subset
of $\mathcal{K}_j$ equaling to ${\cal T}_j$.

\vskip 3mm \par
 For the structure and properties of $\mathcal{K}_j$, we have the following lemma.

\vskip 3mm
\noindent
  {\bf Lemma 3.1} (cf. [8] Example 2.1) \textit{Let $1\leq j\leq r$. Then we have the following conclusions}.

\vskip 2mm\par
  (i) \textit{$\mathcal{K}_j$ is a finite chain ring, $\pi_j$ generates the unique
maximal ideal $\langle \pi_j\rangle=\pi_j\mathcal{K}_j$ in $\mathcal{K}_j$, the nilpotency index of $\pi_j$ is equal to $\lambda p^k$}.
\textit{$\mathcal{K}_j/\langle \pi_j\rangle$ is the residue class field of $\mathcal{K}_j$
modulo $\langle \pi_j\rangle$ and $\mathcal{K}_j/\langle \pi_j\rangle\cong \mathcal{F}_j$}.

\vskip 2mm\par
  (ii) \textit{Every element $\xi$ of $\mathcal{K}_j$ has a unique $\pi_j$-adic expansion:
$$\xi=\sum_{s=0}^{\lambda p^k-1}\pi_j^sb_s(x), \
{\rm where} \ b_s(x)\in {\cal T}_j, \ s=0,1,\ldots,\lambda p^k-1.$$
 Hence $|\mathcal{K}_j|=|{\cal T}_j|^{\lambda p^k}=p^{md_j
\lambda p^k}$. Moreover, $\xi\in \mathcal{K}_j^{\times}$ if and only if $b_0(x)\neq 0$}.

\vskip 2mm\par
  (iii) \textit{All distinct ideals of $\mathcal{K}_j$ are given by:
$\langle \pi_j^l\rangle=\pi_j^l\mathcal{K}_j, \ 0\leq l\leq \lambda p^k.$
 Let
$\mathcal{K}_j/\langle \pi_j^l\rangle$ be the residue class ring of $\mathcal{K}_j$
modulo $\langle \pi_j^l\rangle$. Then
$\mathcal{K}_j/\langle \pi_j^l\rangle\cong \mathbb{F}_{p^m}[x]/\langle \pi_j^l\rangle$ as rings when $l\geq 1$. We write $\mathcal{K}_j/\langle \pi_j^0\rangle=\{0\}$}

\vskip 2mm\par
  (iv) \textit{Let $1\leq l\leq \lambda p^k$. We can
identify $\mathcal{K}_j/\langle \pi_j^l\rangle$
with
$$\mathbb{F}_{p^m}[x]/\langle \pi_j^l\rangle=\{\sum_{i=0}^{l-1}b_i(x)\pi_j^i\mid b_i(x)\in \mathcal{T}_j, \ 0\leq i\leq l-1\}
 \ (\pi_j^l=0)$$
as sets.
Then $\mathcal{K}_j/\langle \pi_j^l\rangle$ is a finite chain ring, $\pi_j(\mathcal{K}_j/\langle \pi_j^l\rangle)$ is the unique maximal ideal in $\mathcal{K}_j/\langle \pi_j^l\rangle$ and the nilpotency index of $\pi_j$ is equal to $l$.
All distinct ideals
of $\mathcal{K}_j/\langle \pi_j^l\rangle$ are given by: $\pi_j^s(\mathcal{K}_j/\langle\pi_j^l\rangle)$, $0\leq s\leq l$}.

\vskip 2mm\par
  (v) \textit{For any $1\leq l\leq \lambda p^k-1$, we have
$$\pi_j(\mathcal{K}_j/\langle\pi_j^l\rangle)=\{\sum_{i=1}^{l-1}b_i(x)\pi_j^i\mid
b_1(x),\ldots,b_{l-1}(x)\in\mathcal{T}_j\}  \ (\pi_j^l=0).$$
Hence $|\pi_j(\mathcal{K}_j/\langle\pi_j^l\rangle)|=|\mathcal{T}_j|^{l-1}=p^{md_j(l-1)}$. We set $\pi_j(\mathcal{K}_j/\langle\pi_j\rangle)=\{0\}$ for convenience}.

\vskip 3mm\par
  Using the notations of Lemma 2.5, the operations on $\mathcal{K}_j+u\mathcal{K}_j$ $(u^2=\omega_j\pi_j^{p^k})$ are defined by

\par
  $\diamond$ $(\xi_0+u\xi_1)+(\eta_0+u\eta_1)=(\xi_0+\eta_0)+u(\xi_1+\eta_1)$,

\par
  $\diamond$ $(\xi_0+u\xi_1)(\eta_0+u\eta_1)=\left(\xi_0\eta_0+\omega_j\pi_j^{p^k}\xi_1\eta_1\right)+u(\xi_0\eta_1+\xi_1\eta_0)$,

\noindent
  for all $\xi_0,\xi_1,\eta_0,\eta_1\in \mathcal{K}_j$. Obviously, $\mathcal{K}_j$ is a subring of
$\mathcal{K}_j+u\mathcal{K}_j$.

\par
    Now, we consider how to determine all ideals of the ring $\mathcal{K}_j+u\mathcal{K}_j$. Since $\mathcal{K}_j$ is a subring of
$\mathcal{K}_j+u\mathcal{K}_j$, we see that $\mathcal{K}_j+u\mathcal{K}_j$ is a free $\mathcal{K}_j$-module
of rank $2$ with the basis $\{1,u\}$. Now, we define
$$\theta: \mathcal{K}_j^2\rightarrow \mathcal{K}_j+u\mathcal{K}_j
\ {\rm via} \ (a_0,a_1)\mapsto a_0+ua_1 \ (\forall a_0,a_1\in \mathcal{K}_j).$$
One can easily verify that $\theta$ is an $\mathcal{K}_j$-module isomorphism from $\mathcal{K}_j^2$
onto $\mathcal{K}_j+u\mathcal{K}_j$. The following lemma can be verified by an argument similar to the proof of
Cao et al. [7] Lemma 3.7. Here, we omit its proof.

\vskip 3mm
\noindent
  {\bf Lemma 3.2} \textit{Using the notations above, $C$ is an ideal
of the ring $\mathcal{K}_j+u\mathcal{K}_j$ $(u^2=\omega_j\pi_j^{p^k})$ if and only if
there is a unique $\mathcal{K}_j$-submodule $S$ of $\mathcal{K}_j^2$ satisfying}
\begin{equation}
(\omega_j\pi_j^{p^k}a_1,a_0)\in S, \ \forall (a_0,a_1)\in S
\end{equation}
\textit{such that $C=\theta(S)$}.

\vskip 3mm \par
  Recall that every $\mathcal{K}_j$-submodule of $\mathcal{K}_j^2$ is called a \textit{linear code}
 over the finite chain ring $\mathcal{K}_j$ of length $2$. A general discussion and description for
linear codes over arbitrary finite chain ring can be found in [21].
 Let $S$ be a linear code over $\mathcal{K}_j$ of length $2$. A matrix $G$ is called a \textit{generator matrix} for $S$ if every codeword in $S$
is a $\mathcal{K}_j$-linear combination of the row vectors of $G$ and
any row vector of $G$ can not be written as a $\mathcal{K}_j$-linear combination of the other row vectors of $G$.

\par
  In the following lemma, we use lowercase letters to denote the elements of $\mathcal{K}_j$ and
$\mathcal{K}_j/\langle \pi_j^l\rangle$ ($1\leq l\leq \lambda p^k-1$)
in order to simplify the expressions.

\vskip 3mm \noindent
   {\bf Lemma 3.3}  (cf. [8] Lemma 2.2 and Example 2.5) \textit{Using the notations above,
$\sum_{i=0}^{\lambda p^k}(2i+1)p^{md_j(\lambda p^k-i)}$ is the number of
linear codes over
$\mathcal{K}_j$ of length $2$}.

\par
   \textit{Moreover, every linear code over
$\mathcal{K}_j$ of length $2$ has one and only one of the following matrices $G$ as their generator matrices}:

\vskip 2mm \par
(i) \textit{$G=(1,a)$, $a\in \mathcal{K}_j$}.

\vskip 2mm \par
(ii) \textit{$G=(\pi_j^s,\pi_j^{s}a)$, $a\in \mathcal{K}_j/\langle \pi_j^{\lambda p^k-s}\rangle$, $1\leq s\leq \lambda p^k-1$}.

\vskip 2mm \par
(iii) \textit{$G=(\pi_j b,1)$, $b\in \mathcal{K}_j/\langle \pi_j^{\lambda p^k-1}\rangle$}.

\vskip 2mm \par
(iv) \textit{$G=(\pi_j^{s+1}b,\pi_j^s)$, $b\in \mathcal{K}_j/\langle \pi_j^{\lambda p^k-1-s}\rangle$, $1\leq s\leq \lambda p^k-1$}.

\vskip 2mm \par
  (v) \textit{$G=\left(\begin{array}{cc}\pi_j^s & 0\cr0 & \pi_j^s\end{array}\right)$, $0\leq s\leq \lambda p^k$}.

\vskip 2mm \par
  (vi) \textit{$G=\left(\begin{array}{cc}1 & c\cr
0 &\pi_j^t\end{array}\right)$,  $c\in \mathcal{K}_j/\langle \pi_j^{t}\rangle$, $1\leq t\leq \lambda p^k-1$}.

\vskip 2mm \par
  (vii) \textit{$G=\left(\begin{array}{cc}\pi_j^s & \pi_j^sc\cr
0 &\pi_j^{s+t}\end{array}\right)$,  $c\in \mathcal{K}_j/\langle \pi_j^{t}\rangle$, $1\leq t\leq \lambda p^k-1-s$, $1\leq s\leq \lambda p^k-2$}.

\vskip 2mm \par
    (viii) \textit{$G=\left(\begin{array}{cc}c & 1\cr \pi_j^t & 0\end{array}\right)$, $c\in \pi_j(\mathcal{K}_j/\langle \pi_j^{t}\rangle)$, $1\leq t\leq \lambda p^k-1$}.

\vskip 2mm \par
    (ix) \textit{$G=\left(\begin{array}{cc}\pi_j^sc & \pi_j^s\cr \pi_j^{s+t} & 0\end{array}\right)$, $c\in \pi_j(\mathcal{K}_j/\langle \pi_j^{t}\rangle)$,
$1\leq t\leq \lambda p^k-1-s$, $1\leq s\leq \lambda p^k-2$}.

\vskip 3mm\par
   Let $\beta\in \mathcal{K}_j$ and $\beta\neq 0$. By Lemma 3.1(ii), there is a unique integer $t$, $0\leq t\leq \lambda p^k-1$,
such that $\beta=\pi_j^tw$ for some $w\in\mathcal{K}_j^\times$.
We call $t$ the \textit{$\pi_j$-degree} of $\beta$ and denote it by $\|\beta\|_{\pi_j}=t$. If $\beta=0$,
we write $\|\beta\|_{\pi_j}=\lambda p^k$.  For any vector $(\beta_1,\beta_2)\in \mathcal{K}_j^2$, we define the \textit{$\pi_j$-degree} of $(\beta_1,\beta_2)$
by $$\|(\beta_1,\beta_2)\|_{\pi_j}={\rm min}\{\|\beta_1\|_{\pi_j},\|\beta_2\|_{\pi_j}\}.$$
Now, as a special case of [21] Proposition 3.2 and Theorem 3.5,
we deduce the following lemma.

\vskip 3mm \noindent
   {\bf Lemma 3.4} (cf. [8] Lemma 2.3) \textit{Let $S$ be a nonzero linear code over $\mathcal{K}_j$ of length $2$, and $G$ be a generator matrix of $S$ with row vectors $G_1,\ldots,G_\rho\in \mathcal{K}_j^2\setminus\{0\}$ satisfying
$$\|G_j\|_{\pi_j}=t_i, \ {\rm where} \ 0\leq t_1\leq\ldots\leq t_\rho\leq \lambda p^k-1.$$
Then the number of codewords in $S$ is equal to $|S|=|\mathcal{T}_j|^{\sum_{i=1}^\rho(\lambda p^k-t_i)}
=p^{md_j\sum_{i=1}^\rho(\lambda p^k-t_i)}$}.

\vskip 3mm\par
   For any positive integer $i$, let $\lceil\frac{i}{2}\rceil={\rm min}\{l\in\mathbb{Z}^{+}\mid
l\geq \frac{i}{2}\}$ and $\lfloor\frac{i}{2}\rfloor={\rm max}\{l\in\mathbb{Z}^{+}\cup\{0\}\mid
l\leq \frac{i}{2}\}$. It is well known that $\lceil\frac{i}{2}\rceil+\lfloor\frac{i}{2}\rfloor=i$. Using these notations,
we list all distinct $\mathcal{K}_j$-submodules of $\mathcal{K}_j^2$ satisfying Condition (4) in Lemma 3.2
by the following lemma.

\vskip 3mm \noindent
   {\bf Lemma 3.5} (cf. [9] Theorem 4) \textit{Using the notations above, every linear code $S$ over
$\mathcal{K}_j$ of length $2$ satisfying Condition $(4)$ in Lemma 3.2
has one and only one of the following matrices $G$ as its generator matrix}:

\vskip 2mm \par
    (I) \textit{$G=(0,\pi_j^{\lambda p^k-1})$; $G=(\pi_j^{\lambda p^k-1} b(x),\pi_j^{\lambda p^k-2})$ where $b(x)\in \mathcal{T}_{j}$};

\par
   \textit{$G=(\pi_j^{\lceil \frac{\lambda p^k-s}{2}\rceil+s} h(x), \pi_j^s)$, where $h(x)\in \mathcal{K}_j/\langle \pi_j^{\lfloor\frac{\lambda p^k-s}{2}\rfloor}\rangle$ and
$(\lambda-1)p^k\leq s\leq \lambda p^k-3$}.

\vskip 2mm\par
   (II) \textit{$G=\left(\begin{array}{cc}\pi_j^s & 0\cr
0 & \pi_j^s\end{array}\right)$, $0\leq s\leq \lambda p^k$}.

\vskip 2mm\par
   (III) \textit{$G=\left(\begin{array}{cc}0 & 1\cr
\pi_j & 0\end{array}\right)$; $G=\left(\begin{array}{cc}\pi_j h(x) & 1\cr
\pi_j^2 & 0\end{array}\right)$ where $h(x)\in \mathcal{T}_{j}$};

\par
  \textit{$G=\left(\begin{array}{cc} \pi_j^{\lceil \frac{t}{2}\rceil}h(x) & 1\cr
\pi_j^t & 0\end{array}\right)$ where $h(x)\in \mathcal{K}_j/\langle \pi_j^{\lfloor \frac{t}{2}\rfloor}\rangle$
and $3\leq t\leq p^k$}.

\vskip 2mm\par
   (IV) \textit{$G=\left(\begin{array}{cc}0 & \pi_j^s\cr
\pi_j^{s+1} & 0\end{array}\right)$ where $1\leq s\leq \lambda p^k-2$};

\par
  \textit{$G=\left(\begin{array}{cc} \pi_j^{s+1}h(x) & \pi_j^s\cr
\pi_j^{s+2} & 0\end{array}\right)$ where $h(x)\in\mathcal{T}_{j}$ and $1\leq s\leq \lambda p^k-3$};

\par
 \textit{$G=\left(\begin{array}{cc} \pi_j^{s+\lceil \frac{t}{2}\rceil}h(x) & \pi_j^{s}\cr
\pi_j^{s+t} & 0\end{array}\right)$ where $h(x)\in \mathcal{K}_j/\langle \pi_j^{\lfloor\frac{t}{2}\rfloor}\rangle$, $1\leq s\leq \lambda p^k-1-t$ and $3\leq t\leq p^k$}.

\vskip 3mm\par
   Let $C$ be an ideal of the ring $\mathcal{K}_j+u\mathcal{K}_j$.
The \textit{annihilating ideal} of $C$ is defined by
${\rm Ann}(C)=\{\xi\in \mathcal{K}_j+u\mathcal{K}_j\mid \xi\eta=0, \ \forall \eta\in C\}$. Now, we give an explicit representation
for all distinct ideals of $\mathcal{K}_j+u\mathcal{K}_j$ and their annihilating ideals, where $1\leq j\leq r$.

\vskip 3mm
\noindent
  {\bf Theorem 3.6} \textit{All distinct ideals of the ring $\mathcal{K}_j+u\mathcal{K}_j$
and their annihilating ideals are given by
one the following four cases}:

\vskip 2mm\par
  (I) \textit{$1+p^{md_j}+\sum_{t=3}^{p^k}p^{md_j\lfloor\frac{t}{2}\rfloor}$ ideals}:

\vskip 2mm\par
  (i-1) \textit{$C=\langle u \pi_j^{\lambda p^k-1}\rangle$ with $|C|=p^{md_j}$ and ${\rm Ann}(C)=\langle u,\pi_j\rangle$};

\vskip 2mm\par
  (i-2) \textit{$C=\langle \pi_j^{\lambda p^k-1}b(x)+u\pi_j^{\lambda p^k-2}\rangle$ with $|C|=p^{2md_j}$,
where $b(x)\in \mathcal{T}_{j}$,  and
${\rm Ann}(C)=\langle \pi_j(-b(x))+u,\pi_j^2\rangle$};

\vskip 2mm\par
  (i-3) \textit{$C=\langle \pi_j^{\lceil\frac{\lambda p^k-s}{2}\rceil+s} h(x)+ u \pi_j^{s}\rangle$
with $|C|=p^{md_j(\lambda p^k-s)}$, where $h(x)\in \mathcal{K}_{j}/\langle \pi_j^{\lfloor\frac{\lambda p^k-s}{2}\rfloor}\rangle$
and $(\lambda-1)p^k\leq s\leq \lambda p^k-3$, and}
$${\rm Ann}(C)=\langle \pi_j^{\lceil\frac{\lambda p^k-s}{2}\rceil}(-h(x))+u,\pi_j^{\lambda p^k-s}\rangle.$$

\par
  (II) \textit{$\lambda p^k+1$ ideals}:

\par
   \textit{$C=\langle \pi_j^s\rangle$ with $|C|=p^{2md_j(\lambda p^k-s)}$ and ${\rm Ann}(C)=\langle \pi_j^{\lambda p^k-s}\rangle$,  $0\leq s\leq \lambda p^k$}.

\vskip 2mm\par
  (III) \textit{$1+p^{md_j}+\sum_{t=3}^{p^k}p^{md_j\lfloor\frac{t}{2}\rfloor}$ ideals}:

\vskip 2mm\par
  (iii-1) \textit{$C=\langle u,\pi_j\rangle$ with $|C|=p^{md_j(2\lambda p^k-1)}$ and ${\rm Ann}(C)=\langle u \pi_j^{\lambda p^k-1}\rangle$};

\vskip 2mm\par
  (iii-2) \textit{$C=\langle \pi_jb(x)+u,\pi_j^2\rangle$ with $|C|=p^{2md_j(\lambda p^k-1)}$,
where $b(x)\in \mathcal{T}_{j}$, and
${\rm Ann}(C)=\langle \pi_j^{\lambda p^k-1}(-b(x))+u\pi_j^{\lambda p^k-2}\rangle$};

\vskip 2mm\par
  (iii-3) \textit{$C=\langle \pi_j^{\lceil \frac{t}{2}\rceil}h(x)+u,\pi_j^t\rangle$ with $|C|=p^{md_j(2\lambda p^k-t)}$, where $3\leq t\leq p^k$ and $h(x)\in \mathcal{K}_{j}/\langle \pi_j^{\lfloor \frac{t}{2}\rfloor}\rangle$,  and
${\rm Ann}(C)=\langle \pi_j^{\lambda p^k-t+\lceil\frac{t}{2}\rceil}(-h(x))+ u \pi_j^{\lambda p^k-t}\rangle$}.

\vskip 2mm\par
  (IV) \textit{$\lambda p^k-2+(\lambda p^k-3)p^{md_j}+\sum_{t=3}^{p^k}(\lambda p^k-1-t)p^{md_j\lfloor\frac{t}{2}\rfloor}$ ideals}:

\vskip 2mm\par
  (iv-1) \textit{$C=\langle \pi_j^{s+1}, u\pi_j^s\rangle$ with $|C|=p^{md_j(2\lambda p^k-2s-1)}$,
  where $1\leq s\leq \lambda p^k-2$, and
${\rm Ann}(C)=\langle  \pi_j^{\lambda p^k-s}, u\pi_j^{\lambda p^k-s-1}\rangle$};

\vskip 2mm\par
  (iv-2) \textit{$C=\langle  \pi_j^{s+1}b(x)+u\pi_j^s,  \pi_j^{s+2}\rangle$ with $|C|=p^{2md_j(\lambda p^k-s-1)}$,
where
$b(x)\in \mathcal{T}_{j}$ and $1\leq s\leq \lambda p^k-3$,
and}
 $${\rm Ann}(C)=\langle \pi_j^{\lambda p^k-s-1}(-b(x))+u\pi_j^{\lambda p^k-s-2}, \pi_j^{\lambda p^k-s} \rangle.$$

\par
  (iv-3) \textit{$C=\langle  \pi_j^{s+\lceil \frac{t}{2}\rceil}h(x)+u\pi_j^s, \pi_j^{s+t} \rangle$ with $|C|=p^{md_j(2\lambda p^k-2s-t)}$,
where
$h(x)\in \mathcal{K}_{j}/\langle \pi_j^{\lfloor \frac{t}{2}\rfloor}\rangle$, $1\leq s\leq \lambda p^k-t-1$
and $3\leq t\leq p^k$, and}
$${\rm Ann}(C)=\langle \pi_j^{\lambda p^k-s-t+\lceil \frac{t}{2}\rceil}(-h(x))+u\pi_j^{\lambda p^k-s-t},
\pi_j^{\lambda p^k-s}\rangle.$$

\par
  \textit{Therefore, the number of ideals in $\mathcal{K}_j+u\mathcal{K}_j$ is equal to}
$$N_{(p^m,2\lambda,p^k,d_j)}=\sum_{l=0}^{\frac{p^k-1}{2}}\left(1+2\lambda p^k-4l\right)p^{lmd_j}.$$

\noindent
   {\bf Proof.} Let $C$ be an ideal
of the ring $\mathcal{K}_j+u\mathcal{K}_j$ $(u^2=\omega_j\pi_j^{p^k})$. By Lemma 3.2 there is a unique
$\mathcal{K}_j$-submodule $S$ of $\mathcal{K}_j^2$ satisfying Condition (4) such that $C=\theta(S)$. From this and by
Lemma 3.5, we deduce that $S$ has one and only one of the following matrices $G$ as its generator matrix:

\par
   (i) $G$ is given in Lemma 3.5(I). Hence we have three subcases:

\par
   (i-1) $G=(0,\pi_j^{\lambda p^k-1})$. In this case, we have $C=\theta(S)=\langle \theta(0,\pi_j^{\lambda p^k-1})\rangle
=\langle u \pi_j^{\lambda p^k-1}\rangle$. From $\|(0,\pi_j^{\lambda p^k-1})\|_{\pi_j}=\lambda p^k-1$ and Lemma 3.4, we deduce that $|S|=p^{md_j(\lambda p^k-(\lambda p^k-1))}=p^{md_j}$. Hence
$|C|=|S|=p^{md_j}$, since $\theta$ is a bijection.

\par
   (i-2) $G=(\pi_j^{\lambda p^k-1}b(x),\pi_j^{\lambda p^k-2})$ where $b(x)\in \mathcal{T}_{j}$. In this case, we have
$$C=\theta(S)=\langle \theta(\pi_j^{\lambda p^k-1}b(x),\pi_j^{\lambda p^k-2})\rangle
=\langle \pi_j^{\lambda p^k-1}b(x)+u\pi_j^{\lambda p^k-2}\rangle.$$
Moreover, by $\|(\pi_j^{\lambda p^k-1}b(x),\pi_j^{\lambda p^k-2})\|_{\pi_j}=\lambda p^k-2$ and Lemma 3.4 we deduce that
$|S|=p^{md_j(\lambda p^k-(\lambda p^k-2))}=p^{2md_j}$. This implies
$|C|=|S|=p^{2md_j}$.

\par
   (i-3) $G=(\pi_j^{\lceil\frac{\lambda p^k-s}{2}\rceil+s} h(x), \pi_j^s)$, where $h(x)\in \mathcal{K}_j/\langle \pi_j^{\lfloor\frac{\lambda p^k-s}{2}\rfloor}\rangle$ and
$(\lambda-1)p^k\leq s\leq \lambda p^k-3$. In this case, we have
$C=\theta(S)=\langle \theta(\pi_j^{\lceil\frac{\lambda p^k-s}{2}\rceil+s} h(x), \pi_j^s)\rangle
=\langle \pi_j^{\lceil\frac{\lambda p^k-s}{2}\rceil+s} h(x)+ u \pi_j^{s}\rangle.$

\par
   As $\|(\pi_j^{\lceil\frac{\lambda p^k-s}{2}\rceil+s} h(x),\pi_j^s)\|_{\pi_j}=s$, by Lemma 3.4 we have $|S|=p^{md_j(\lambda p^k-s)}$. Hence
$|C|=|S|=p^{md_j(\lambda p^k-s)}$, since $\theta$ is a bijection.

\par
   Therefore, the number $N_{(I)}$ of ideals in $\mathcal{K}_j+u\mathcal{K}_j$ in Case (I) is equal to
\begin{center}
$N_{(I)}=1+|\mathcal{T}_{j}|+\sum_{s=(\lambda-1)p^k}^{\lambda p^k-3}|\mathcal{T}_{j}|^{\lfloor
\frac{\lambda p^k-s}{2}\rfloor}
=1+p^{md_j}+\sum_{s=(\lambda-1)p^k}^{\lambda p^k-3}p^{md_j\lfloor
\frac{\lambda p^k-s}{2}\rfloor}$.
\end{center}
Now, let $t=\lambda p^k-s$. Then $s=\lambda p^k-t$ where $3\leq t\leq p^k$, and hence
 $N_{(I)}=1+p^{md_j}+\sum_{t=3}^{p^k}p^{md_j\lfloor\frac{t}{2}\rfloor}$.

\par
   (ii) $G=\left(\begin{array}{cc}\pi_j^s & 0\cr
0 & \pi_j^s\end{array}\right)$, where $0\leq s\leq \lambda p^k$. In this case, we have $C=\langle \theta(\pi_j^s,0),\theta(0,\pi_j^s)\rangle=\langle \pi_j^s,u\pi_j^s\rangle=\langle \pi_j^s\rangle.$
By $\|(\pi_j^s,0)\|_{\pi_j}=\|(0,\pi_j^s)\|_{\pi_j}=s$ and Lemma 3.4, we deduce that $|C|=|S|=p^{md_j((\lambda p^k-s)+(\lambda p^k-s))}=p^{2md_j(\lambda p^k-s)}$.

\par
  (iii) Similar to the case (i), we have three subcases:

\par
  (iii-1) $G=\left(\begin{array}{cc}0 & 1\cr
\pi_j & 0\end{array}\right)$. In this case, we have $C=\langle \theta(0,1),\theta(\pi_j,0)\rangle=\langle u,\pi_j\rangle$.
Then by Lemma 3.4, $\|(0,1)\|_{\pi_j}=0$ and $\|(\pi_j,0)\|_{\pi_j}=1$ we have $|C|=|S|=p^{md_j((\lambda p^k-0)+(\lambda p^k-1))}=p^{md_j(2\lambda p^k-1)}$.

\par
  (iii-2) $G=\left(\begin{array}{cc}\pi_jh(x) & 1\cr
\pi_j^2 & 0\end{array}\right)$ where $h(x)\in \mathcal{T}_{j}$. In this case, we have
$C=\langle \theta(\pi_jh(x),1),\theta(\pi_j^2,0)\rangle
=\langle \pi_jh(x)+u,\pi_j^2\rangle$.
By Lemma 3.4, $\|(\pi_jh(x),1)\|_{\pi_j}=0$ and $\|(\pi_j^2,0)\|_{\pi_j}=2$ we have $|C|=|S|=p^{md_j((\lambda p^k-0)+(\lambda p^k-2))}=p^{2md_j(\lambda p^k-1)}$.

\par
  (iii-3) $G=\left(\begin{array}{cc} \pi_j^{\lceil \frac{t}{2}\rceil}h(x) & 1\cr
\pi_j^t & 0\end{array}\right)$ where $h(x)\in \mathcal{K}_j/\langle \pi_j^{\lfloor\frac{t}{2}\rfloor}\rangle$
and $3\leq t\leq p^k$. In this case, we have
$C=\langle \theta(\pi_j^{\lceil \frac{t}{2}\rceil}h(x),1),\theta(\pi_j^t,0)\rangle
=\langle \pi_j^{\lceil \frac{t}{2}\rceil}h(x)+u,\pi_j^t\rangle$.

\par
   Moreover, by $\|(\pi_j^{\lceil \frac{t}{2}\rceil}h(x),1)\|_{\pi_j}=0$, $\|(\pi_j^t,0)\|_{\pi_j}=t$ and Lemma 3.4, we deduce that $|C|=|S|=p^{md_j((\lambda p^k-0)+(\lambda p^k-t))}=p^{md_j(2\lambda p^k-t)}$.

\par
   Therefore, the number $N_{(III)}$ of ideals of $\mathcal{K}_j+u\mathcal{K}_j$ in Case (III) is equal to $N_{(III)}=1+|\mathcal{T}_{j}|+\sum_{t=3}^{p^k}|\mathcal{T}_{j}|^{\lfloor \frac{t}{2}\rfloor}
=1+p^{md_j}+\sum_{t=3}^{p^k}p^{md_j\lfloor \frac{t}{2}\rfloor}.$

\par
  (iv) Similar to the cases (i) and (iii), we have three subcases:

\par
   (iv-1) $G=\left(\begin{array}{cc}0 & \pi_j^s\cr
\pi_j^{s+1} & 0\end{array}\right)$, where $1\leq s\leq \lambda p^k-2$. In this case, we have $C=\langle \theta(0,\pi_j^s),\theta(\pi_j^{s+1},0)\rangle=\langle u\pi_j^s,\pi_j^{s+1}\rangle$.

\par
   By $\|(0,\pi_j^s)\|_{\pi_j}=s$, $\|(\pi_j^{s+1},0)\|_{\pi_j}=s+1$ and Lemma 3.4, we obtain $|C|=|S|=p^{md_j((\lambda p^k-s)+(\lambda p^k-(s+1)))}=p^{md_j(2\lambda p^k-2s-1)}$.

\par
   (iv-2) $G=\left(\begin{array}{cc} \pi_j^{s+1}b(x) & \pi_j^s\cr
\pi_j^{s+2} & 0\end{array}\right)$, where $b(x)\in\mathcal{T}_{j}$ and $1\leq s\leq \lambda p^k-3$.
Then $C=\langle \theta(\pi_j^{s+1}b(x),\pi_j^s),\theta(\pi_j^{s+2},0)\rangle=\langle \pi_j^{s+1}b(x)+u\pi_j^s,\pi_j^{s+2}\rangle$.

\par
   By $\|(\pi_j^{s+1}b(x),\pi_j^s)\|_{\pi_j}=s$, $\|(\pi_j^{s+2},0)\|_{\pi_j}=s+2$ and Lemma 3.4, we deduce that $|C|=|S|=p^{md_j((\lambda p^k-s)+(\lambda p^k-(s+2)))}=p^{2md_j(\lambda p^k-s-1)}$.

\par
   (iv-3) $G=\left(\begin{array}{cc} \pi_j^{s+\lceil \frac{t}{2}\rceil}h(x) & y^{s}\cr
\pi_j^{s+t} & 0\end{array}\right)$, where $h(x)\in \mathcal{K}_j/\langle \pi_j^{\lfloor \frac{t}{2}\rfloor}\rangle$,
$1\leq s\leq \lambda p^k-1-t$ and $3\leq t\leq p^k$.
In this case, we have
$$C=\langle \theta(\pi_j^{s+\lceil \frac{t}{2}\rceil}h(x),\pi_j^s),\theta(\pi_j^{s+t},0)\rangle=\langle \pi_j^{s+\lceil \frac{t}{2}\rceil}h(x)+u\pi_j^s,\pi_j^{s+t}\rangle.$$

\par
   By $\|(\pi_j^{s+\lceil \frac{t}{2}\rceil}h(x),\pi_j^s)\|_{\pi_j}=s$, $\|(\pi_j^{s+t},0)\|_{\pi_j}=s+t$ and Lemma 3.4, we deduce that $|C|=|S|=p^{md_j((\lambda p^k-s)+(\lambda p^k-(s+t)))}=p^{md_j(2\lambda p^k-2s-t)}$.

\par
   Hence the number $N_{(IV)}$ of ideals of $\mathcal{K}_j+u\mathcal{K}_j$ in Case (IV) is equal to
$N_{(IV)}=\lambda p^k-2+(\lambda p^k-3)p^{md_j}+\sum_{t=3}^{p^k}(\lambda p^k-1-t)p^{md_j\lfloor\frac{t}{2}\rfloor}.$

\par
   As stated above, we see that the number of ideals in $\mathcal{K}_j+u\mathcal{K}_j$ is equal to
\begin{eqnarray*}
N_{(p^m,2\lambda,p^k,d_j)}&=&N_{(I)}+\lambda p^k+1+N_{(III)}+N_{(IV)}\\
 &=&1+2\lambda p^k+(\lambda p^k-1)p^{md_j}+\sum_{t=3}^{p^k}(\lambda p^k-t+1)p^{md_j\lfloor\frac{t}{2}\rfloor}\\
 &=&\sum_{l=0}^{\frac{p^k-1}{2}}(1+2\lambda p^k-4l)p^{lmd_j}.
\end{eqnarray*}

\par
  Since $\mathcal{K}_j$ is a finite chain ring which is a special Frobenius ring, $\pi_j^{\lambda p^k}=0$ in $\mathcal{K}_j$ and $u^2=\omega_j\pi_j^{p^k}$ in $\mathcal{K}_j+u\mathcal{K}_j$, one can
easily verify the conclusions for the annihilating ideal ${\rm Ann}(C)$ of each ideal $C$ listed in this theorem by
the known results on linear codes over commutative Frobenius rings
(cf. [18]). Here, we
omit the proofs.
\hfill $\Box$

\vskip 3mm\noindent
  {\bf Remark} By Theorem 3.6, for any ideal $C_j$ of $\mathcal{K}_j+u\mathcal{K}_j$ it can be verified by a direct calculation that
\begin{equation}
|C_j||{\rm Ann}(C_j)|=p^{2md_j\lambda p^k}.
\end{equation}

\par
   From Theorems 2.7 and 3.6, we deduce the following conclusion.

\vskip 3mm\noindent
  {\bf Corollary 3.7} \textit{Every $(\delta+\alpha u^2)$-constacyclic code ${\cal C}$ over $R$ of length $np^k$
can be constructed by the following two steps}:

\vskip 2mm\par
  (i) \textit{For each integer $j=1,\ldots,r$, choose an ideal $C_j$ of
${\cal K}_j+u{\cal K}_j$ listed in Theorem 3.6}.

\vskip 2mm\par
  (ii) \textit{Set $\mathcal{C}=\sum_{j=1}^r e_j(x)\Psi(C_j)$} (mod $x^{np^k}-(\delta+\alpha u^2)$).
\textit{The number of codewords in ${\cal C}$ is
equal to $|{\cal C}|=\prod_{j=0}^r|C_j|$}.

\vskip 2mm\par
  \textit{Therefore, the number $\mathcal{N}$ of $(\delta+\alpha u^2)$-constacyclic code ${\cal C}$ over $R$ of length $np^k$ is equal to $\mathcal{N}=\prod_{j=1}^rN_{(p^m,2\lambda,p^k,d_j)}
  = \prod_{j=1}^r(\sum_{l=0}^{\frac{p^k-1}{2}}(1+2\lambda p^k-4l)p^{lmd_j})$}.

\vskip 3mm\noindent
  {\bf Remark} Let $p=3$, $k=1$ and $\lambda=2$. There is a unique integer $s=3$ satisfying
$(\lambda-1)p^k\leq s\leq \lambda p^k-3$ in Case (i-3) of Theorem 3.6,
and there is a unique integer $t=3$ satisfying
$3\leq t\leq p^k$ in Cases (iii-3) and (iv-3)  of Theorem 3.6. For this special case, a complete description for $(\delta+\alpha u^2)$-constacyclic
codes over $\mathbb{F}_{3^m}[u]/\langle u^{4}\rangle$ of length
$3n$ had been given (cf. [6]), where ${\rm gcd}(3,n)=1$.
It is clear that Theorem 3.6 is a nontrivial promotion for that in [6].

\vskip 3mm\par
   Using the notations of Corollary 3.7, ${\cal C}=\bigoplus_{j=1}^re_j(x)\Psi(C_j)$ is called the \textit{canonical form decomposition} of the $(\delta+\alpha u^2)$-constacyclic code $\mathcal{C}$.  Finally, we give
the following conclusion.

\vskip 3mm \noindent
  {\bf Corollary 3.8} \textit{Using the notations of Corollary 3.7, for each integer $j$, $1\leq j\leq r$, there exist
$\beta_{j1}(x),\beta_{j2}(x)\in \mathcal{K}_j+u\mathcal{K}_j$ such that
$C_j=\langle \beta_{j1}(x),\beta_{j2}(x)\rangle$ where we set $\beta_{j2}(x)=0$ if $C_j$ is principal. Then}
$$\mathcal{C}=\bigoplus_{j=1}^re_j(x)\Psi(C_j)=\left\langle \vartheta_1(x), \vartheta_2(x)\right\rangle,$$
\textit{where $\vartheta_s(x)=\sum_{j=1}^re_j(x)\Psi(\beta_{js}(x))\in R[x]/\langle x^{np^k}-(\delta+\alpha u^2)\rangle$ for $s=1,2$}.

\vskip 3mm \noindent
  {\bf Proof.} By Lemma 2.5 it follows that
$$\mathcal{A}+u\mathcal{A}=\{\sum_{j=1}^r\varepsilon_j(x)\xi_j\mid
\xi_j\in \mathcal{K}_j+u\mathcal{K}_j, \ j=1,\ldots,r\}.$$
Then by Theorem 2.1, $R[x]/\langle x^{np^k}-(\delta+\alpha u^2)\rangle
 =\Psi(\mathcal{A}+u\mathcal{A})$ and $e_j(x)=\Psi(\varepsilon_j(x))$ for all $j$, we obtain
$$R[x]/\langle x^{np^k}-(\delta+\alpha u^2)\rangle
=\{\sum_{j=1}^re_j(x)\Psi(\xi_j)\mid
\xi_j\in \mathcal{K}_j+u\mathcal{K}_j, \ j=1,\ldots,r\}.$$
Now, let $1\leq j\leq r$. Since $C_j$ is an ideal of the ring $\mathcal{K}_j+u\mathcal{K}_j$
generated by $\beta_{j1}(x)$ and $\beta_{j2}(x)$, we have $C_j=\{\xi_{j1}\beta_{j1}(x)+\xi_{j2}\beta_{j2}(x)\mid \xi_{j1},\xi_{j2}\in
\mathcal{K}_j+u\mathcal{K}_j\}$.
As $e_j(x)^2=e_j(x)$ and $e_j(x)e_l(x)=0$ for all $1\leq j\neq l\leq r$ by Lemma 2.6(i),
from Theorem 2.1 we deduce that
\begin{eqnarray*}
\mathcal{C}&=&\sum_{j=1}^r\{\sum_{s=1,2}e_j(x)\Psi(\xi_{js})\Psi(\beta_{js}(x))\mid \xi_{j1},\xi_{j2}\in
\mathcal{K}_j+u\mathcal{K}_j\}\\
 &=&\{\sum_{s=1,2}(\sum_{j=1}^re_j(x)\Psi(\xi_{js}))\cdot\vartheta_{s}(x)
  \mid \xi_{j1},\xi_{j2}\in
\mathcal{K}_j+u\mathcal{K}_j, 1\leq j\leq r\}\\
  &=&\{\eta_1(x)\vartheta_{1}(x)+\eta_2(x)\vartheta_{2}(x)\mid
  \eta_1(x),\eta_2(x)\in R[x]/\langle x^{np^k}-(\delta+\alpha u^2)\rangle\}\\
  &=&\langle \vartheta_{1}(x),\vartheta_{2}(x)\rangle.
\end{eqnarray*}
Therefore, $\mathcal{C}$ is an ideal of $R[x]/\langle x^{np^k}-(\delta+\alpha u^2)\rangle$
generated by $\vartheta_{1}(x)$ and $\vartheta_{2}(x)$.
\hfill $\Box$

\vskip 3mm \par
  Using the notations of Corollary 3.8, we call $\mathcal{C}$ a \textit{1-generator code}
if $\vartheta_{2}(x)=0$, i.e. $\mathcal{C}=\langle \vartheta_{1}(x)\rangle$. Otherwise,
we call $\mathcal{C}$ a \textit{2-generator code}. Then from Theorem 3.6 and Corollary 3.8,
we deduce the following conclusion.

\vskip 3mm \noindent
  {\bf Corollary 3.9} \textit{Using the notations of Corollary 3,7, let $\mathcal{N}_s$ be the number of $s$-generator $(\delta+\alpha u^2)$-constacyclic codes over $R$ of length $np^k$ for $s=1,2$. Then}
$\mathcal{N}_1=\prod_{j=1}^r\left(2+\lambda p^k+p^{md_j}+\sum_{t=3}^{p^k}p^{md_j\lfloor\frac{t}{2}\rfloor}\right)
\ {\rm and} \ \mathcal{N}_2=\mathcal{N}-\mathcal{N}_1.$


\section{Dual codes of $(\delta+\alpha u^2)$-constacyclic codes over $R$} \label{}
\noindent
   In this section, we determine the dual code for every $(\delta+\alpha u^2)$-constacyclic code over $R$ of length $np^k$.
Obviously, we have
$$(\delta+\alpha u^2)^{-1}=\delta^{-1}-\delta^{-2}\alpha u^2+(-1)^2\delta^{-3}\alpha^2 u^4+\ldots
+(-1)^{\lambda-1}\delta^{-\lambda}\alpha^{\lambda-1} u^{2\lambda-2}$$
which is also a unit in $R$ of Type 2. Let $\mathcal{C}$ be a $(\delta+\alpha u^2)$-constacyclic code over $R$ of length $np^k$.
The dual code $\mathcal{C}^{\bot}$ of $\mathcal{C}$ is a $(\delta+\alpha u^2)^{-1}$-constacyclic code $\mathcal{C}$ over $R$ of length $np^k$, i.e. $\mathcal{C}^{\bot}$ is an ideal of $R[x]/\langle x^{np^k}-(\delta+\alpha u^2)^{-1}\rangle$. Hence $\mathcal{C}^{\bot}$ is a constacyclic code over $R$ of Type $2$ as well.

\par
   In the ring $R[x]/\langle x^{np^k}-(\delta+\alpha u^2)^{-1}\rangle$, we have $x^{np^k}=(\delta+\alpha u^2)^{-1}$, i.e.
$(\delta+\alpha u^2)x^{np^k}=1$. The latter implies
\begin{equation}
x^{-1}=(\delta+\alpha u^2)x^{np^k-1} \ {\rm in} \ R[x]/\langle x^{np^k}-(\delta+\alpha u^2)^{-1}\rangle.
\end{equation}
Define a map $\tau:R[x]/\langle x^{np^k}-(\delta+\alpha u^2)\rangle\rightarrow R[x]/\langle x^{np^k}-(\delta+\alpha u^2)^{-1}\rangle$ by $$\tau(a(x))=a(x^{-1})=\sum_{i=0}^{np^k-1}a_ix^{-i}=a_0+(\delta+\alpha u^2)\sum_{i=1}^{np^k-1}a_ix^{np^k-i},$$
for all $a(x)=\sum_{i=0}^{np^k-1}a_ix^{i}\in R[x]/\langle x^{np^k}-(\delta+\alpha u^2)\rangle$ with $a_0,a_1,\ldots,a_{np^k-1}\in R$.
Especially, we have $\tau(x)=(\delta+\alpha u^2)x^{np^k-1}$ by (6).
Then $\tau$ is an isomorphism of rings from $R[x]/\langle x^{np^k}-(\delta+\alpha u^2)\rangle$ onto $R[x]/\langle x^{np^k}-(\delta+\alpha u^2)^{-1}\rangle$.

\par
  Let $\mathcal{C}$ be an ideal of $R[x]/\langle x^{np^k}-(\delta+\alpha u^2)\rangle$. Define
$\tau(\mathcal{C})=\{\tau(\xi)\mid \xi\in \mathcal{C}\}$. Then $\tau(\mathcal{C})$ is an ideal of
$R[x]/\langle x^{np^k}-(\delta+\alpha u^2)^{-1}\rangle$, and the map:
$$\mathcal{C}\mapsto \tau(\mathcal{C})$$
is a bijection from the set of $(\delta+\alpha u^2)$-constacyclic codes over $R$ of length $np^k$
onto the set of $(\delta+\alpha u^2)^{-1}$-constacyclic codes over $R$ of length $np^k$.

\par
  Let $g(x)$ be a polynomial in $R[x]$ of degree $s$, its reciprocal polynomial $x^sg(x^{-1})$
will denoted by $g^\ast(x)$. For
example, if $g(x)=a_0+a_1x+\ldots+a_sx^s$, then $g^\ast(x)=a_s+a_{s-1}x+\ldots+a_0x^s$.

\par
  Let
$\mathcal{C}$ be an ideal of $R[x]/\langle x^{np^k}-(\delta+\alpha u^2)\rangle$. Define a set by
$${\rm Ann}(\mathcal{C})=\{h(x)\in R[x]/\langle x^{np^k}-(\delta+\alpha u^2)\rangle\mid
g(x)h(x)=0, \ \forall g(x)\in \mathcal{C}\}.$$
Then ${\rm Ann}(\mathcal{C})$ is also an ideal of $R[x]/\langle x^{np^k}-(\delta+\alpha u^2)\rangle$
and called the \textit{annihilating ideal} of $\mathcal{C}$. It is known that
the dual code $\mathcal{C}^{\bot}$ of $\mathcal{C}$ is given by
\begin{equation}
\mathcal{C}^{\bot}={\rm Ann}^\ast(\mathcal{C})=\{h^\ast(x)\mid h(x)\in {\rm Ann}(\mathcal{C})\}
\ ({\rm cf}. \ [10] \ {\rm Proposition} \ 2.7).
\end{equation}
It is obvious that $|{\rm Ann}^\ast(\mathcal{C})|=
|{\rm Ann}(\mathcal{C})|=|\tau({\rm Ann}(\mathcal{C}))|$. For any $0\neq h(x)\in {\rm Ann}(\mathcal{C})$
of degree $s\leq np^k-1$, it follows that $h^\ast(x)=x^sh(x^{-1})=x^s\tau(h(x))$ and $\tau(h(x))\in \tau({\rm Ann}(\mathcal{C}))$.
From this we deduce that $h^\ast(x)\in \tau({\rm Ann}(\mathcal{C}))$,
since $\tau({\rm Ann}(\mathcal{C}))$ is an ideal of $R[x]/\langle x^{np^k}-(\delta+\alpha u^2)^{-1}\rangle$. Therefore, ${\rm Ann}^\ast(\mathcal{C})
\subseteq \tau({\rm Ann}(\mathcal{C}))$, and hence ${\rm Ann}^\ast(\mathcal{C})
=\tau({\rm Ann}(\mathcal{C}))$.

\par
  Then from Equation (7), we deduce the following conclusion.

\vskip 3mm\noindent
  {\bf Lemma 4.1} \textit{For any $(\delta+\alpha u^2)$-constacyclic code $\mathcal{C}$ over $R$ of length $np^k$,
the dual codes $\mathcal{C}^{\bot}$ of $\mathcal{C}$ is equal to $\tau({\rm Ann}(\mathcal{C}))$}.

\vskip 3mm \par
  Now, let $\mathcal{C}$ be a $(\delta+\alpha u^2)$-constacyclic code over $R$ of length $np^k$
with canonical form decomposition ${\cal C}=\bigoplus_{j=1}^re_j(x)\Psi(C_j)$, where
$C_j$ is an ideal of the ring $\mathcal{K}_j+u\mathcal{K}_j$ determined by Theorem 3.6
for all $j=1,\ldots,r$.
Denote ${\cal D}=\bigoplus_{j=1}^re_j(x)\Psi({\rm Ann}(C_j))$, where ${\rm Ann}(C_j)$ is the
annihilating ideal of $C_j$ in $\mathcal{K}_j+u\mathcal{K}_j$ determined by Theorem 3.6.
Since
$\Psi$ is a ring isomorphism and $e(x)^2=e_j(x)$ and $e_j(x)e_j(x)=0$
for all $1\leq j\neq l\leq r$, it follows that
\begin{eqnarray*}
\mathcal{C}\cdot \mathcal{D}&=&\left(\sum_{j=1}^re_j(x)\Psi(C_j)\right)\left(\sum_{j=1}^re_j(x)\Psi({\rm Ann}(C_j))\right)\\
 &=&\sum_{j=1}^re_j(x)\Psi\left(C_j\cdot {\rm Ann}(C_j)\right)=\{0\}.
\end{eqnarray*}
On the other hand, from Theorem 3.6, Corollary 3.7, Equation (5) and
$|R|=p^{2m\lambda}$ we deduce that
\begin{eqnarray*}
|\mathcal{C}||\mathcal{D}|&=&\left(\prod_{j=1}^r|C_j|\right)\left(\prod_{j=1}^r|{\rm Ann}(C_j)|\right)
=\prod_{j=1}^r(|C_j|)|{\rm Ann}(C_j)|\\
&=&p^{2m\lambda p^k\sum_{j=1}^rd_j}=p^{2m\lambda p^kn}=|R|^{np^k}\\
&=&|R[x]/\langle x^{np^k}-(\delta+\alpha u^2)\rangle|.
\end{eqnarray*}

\par
  As stated above, we conclude that ${\rm Ann}(\mathcal{C})={\cal D}=\bigoplus_{j=1}^re_j(x)\Psi({\rm Ann}(C_j))$.
Since $\tau$ is an isomorphism of rings, by Lemma 4.1 and $\tau(e_j(x))=e(x^{-1})\in
R[x]/\langle x^{np^k}-(\delta+\alpha u^2)^{-1}\rangle$ we conclude the following theorem.

\vskip 3mm\noindent
  {\bf Theorem 4.2} \textit{Let $\mathcal{C}$ be a $(\delta+\alpha u^2)$-constacyclic code over $R$ of length $np^k$
with canonical form decomposition ${\cal C}=\bigoplus_{j=1}^re_j(x)\Psi(C_j)$, where
$C_j$ is an ideal of the ring $\mathcal{K}_j+u\mathcal{K}_j$ determined by Theorem 3.6
for all $j=1,\ldots,r$. Let ${\rm Ann}(C_j)$ be the
annihilating ideal of $C_j$ in $\mathcal{K}_j+u\mathcal{K}_j$ determined by Theorem 3.6.
Then the dual code of $\mathcal{C}$ is given by}
$$\mathcal{C}^{\bot}=\bigoplus_{j=1}^re_j(x^{-1})\tau\left(\Psi({\rm Ann}(C_j))\right).$$


\section{A subclass of $(\delta+\alpha u^2)$-constacyclic codes over $R$} \label{}
\noindent
In this section, let $x^n-\delta_0$ be an irreducible polynomial in
$\mathbb{F}_{p^m}[x]$ and $\delta=\delta_0^{p^k}$. We consider $(\delta+\alpha u^2)$-constacyclic codes over $R$
of length $np^k$.
In this case, we have $r=1$, $\pi_1=f_1(x)=x^n-\delta_0$, $e_1(x)=1$, $d_1={\rm deg}(x^n-\delta_0)=n$ and

\vskip 2mm\par
  $\bullet$ $\mathcal{K}_1=\mathcal{A}=\mathbb{F}_{p^m}[x]/\langle (x^{np^k}-\delta)^\lambda\rangle
  =\mathbb{F}_{p^m}[x]/\langle (x^{n}-\delta_0)^{\lambda p^k}\rangle$;

\vskip 2mm\par
  $\bullet$ $\mathcal{T}_1=\mathcal{T}=\{\sum_{i=0}^{n-1}a_ix^i\mid a_0,a_1,\ldots,a_{n-1}\in \mathbb{F}_{p^m}\}
\ {\rm with} \ |\mathcal{T}|=p^{mn}$;

\vskip 2mm\par
  $\bullet$ $\mathcal{A}/\langle (x^n-\delta_0)^s\rangle=\{\sum_{i=0}^{s-1}(x^n-\delta_0)^it_i(x)\mid
t_i(x)\in\mathcal{T}, 0\leq i\leq s-1\}$, for any $1\leq s<\lambda p^k$.

\vskip 2mm\par
  As $(x^n-\delta_0)^{p^k}=x^{np^k}-\delta=\alpha u^2$ in $R[x]/\langle x^{np^k}-(\delta+\alpha u^2)\rangle$,
for any integers $l_0$ and $l_1$, where $0\leq l_0\leq p^k-1$ and $0\leq l_1\leq \lambda-1$, it follows that
\begin{equation}
\Psi((x^n-\delta_0)^{l_0+l_1p^k})=(x^n-\delta_0)^{l_0}((x^n-\delta_0)^{p^k})^{l_1}=\alpha^{l_1}(x^n-\delta_0)^{l_0}u^{2l_1}.
\end{equation}
Then for any integer $i$, $1\leq i<p^k$, by $\lambda p^k-i=(p^k-i)+(\lambda-1)p^k$ we have
$$\Psi((x^n-\delta_0)^{\lambda p^k-i})=(x^n-\delta_0)^{p^k-i}(\alpha u^2)^{\lambda-1}
=\alpha^{\lambda-1}(x^n-\delta_0)^{p^k-i} u^{2\lambda-2}.$$

\par
  Now, by Theorem 3.6, Corollary 3.7 and Equation (1), we have the following conclusion.

\vskip 3mm
\noindent
  {\bf Theorem 5.1} \textit{Let $x^n-\delta_0$ be an irreducible polynomial in
$\mathbb{F}_{p^m}[x]$ and set $\delta=\delta_0^{p^k}$. Denote
$$\mathcal{H}_s=\mathcal{A}/\langle (x^n-\delta_0)^s\rangle \
{\rm with} \ |\mathcal{H}_s|=p^{smn}, \
1\leq s<\lambda p^k.$$
Then all $N_{(p^m,2\lambda,p^k,n)}$ distinct nonzero
$(\delta+\alpha u^2)$-constacyclic codes over $R$ of length $np^k$ are given by
one the following two cases}:

\vskip 2mm\noindent
  $\diamondsuit$ \textit{$1+\lambda p^k+p^{mn}+\sum_{t=3}^{p^k}p^{mn\lfloor\frac{t}{2}\rfloor}$ 1-generator codes}:
$$\left\langle h(x)\cdot (x^n-\delta_0)^{i_1}u^{s_1}+(x^n-\delta_0)^{i_2}u^{s_2}\right\rangle$$
\textit{where the pairs $(i_1,s_1)$ and $(i_2,s_2)$ of integers, $h(x)$ and the number $|\mathcal{C}|$
of codewords in $\mathcal{C}$ are given by one of the following four subcases}:
\vskip 2mm\par
  (i-1) \textit{$(i_2,s_2)=(p^k-1,2\lambda-1)$, $h(x)=0$ and $|\mathcal{C}|=p^{mn}$}.

\vskip 2mm\par
  (i-2) \textit{$(i_1,s_1)=(p^k-1,2\lambda-2)$, $(i_2,s_2)=(p^k-2,2\lambda-1)$, $h(x)\in\mathcal{H}_1$ and $|\mathcal{C}|=p^{2mn}$}.

\vskip 2mm\par
  (i-3)  \textit{$(i_1,s_1)=(\lceil\frac{p^k-l_0}{2}\rceil+l_0, 2\lambda-2)$,
$(i_2,s_2)=(l_0,2\lambda-1)$, $h(x)\in\mathcal{H}_{\lfloor\frac{p^k-l_0}{2}\rfloor}$ and $|\mathcal{C}|=p^{mn(p^k-l_0)}$,
where $0\leq l_0\leq p^k-3$}.

\vskip 2mm\par
  (ii) \textit{$(i_2,s_2)=(l_0,2l_1)$, $h(x)=0$ and $|\mathcal{C}|=p^{2mn((\lambda-l_1) p^k-l_0)}$,
where $0\leq l_0\leq p^k-1$
and $0\leq l_1\leq \lambda-1$}.

\vskip 2mm\noindent
  $\diamondsuit\diamondsuit$ \textit{$\lambda p^k-1+(\lambda p^k-2)p^{mn}+\sum_{t=3}^{p^k}(\lambda p^k-t)p^{mn\lfloor\frac{t}{2}\rfloor}$ 2-generator codes}:
$$\left\langle h(x)\cdot (x^n-\delta_0)^{i_1}u^{s_1}+(x^n-\delta_0)^{i_2}u^{s_2}, (x^n-\delta_0)^{i_3}u^{s_3}\right\rangle$$
\textit{where the pairs $(i_1,s_1)$, $(i_2,s_2)$ and $(i_3,s_3)$ of integers, $h(x)$ and the number $|\mathcal{C}|$
of codewords in $\mathcal{C}$ are given by one of the following subcases}:

\vskip 2mm\par
  (iii-1) \textit{$(i_2,s_2)=(0,1)$, $(i_3,s_3)=(1,0)$,
$h(x)=0$ and  $|\mathcal{C}|=p^{mn(2\lambda p^k-1)}$}.

\vskip 2mm\par
  (iii-2) \textit{$(i_1,s_1)=(1,0)$, $(i_2,s_2)=(0,1)$, $(i_3,s_3)=(2,0)$,
$h(x)\in \mathcal{H}_1$ and  $|\mathcal{C}|=p^{2mn(\lambda p^k-1)}$}.

\vskip 2mm\par
  (iii-3-1) \textit{$(i_1,s_1)=(\frac{p^k+1}{2},0)$, $(i_2,s_2)=(0,1)$, $(i_3,s_3)=(0,2)$,
$h(x)\in \mathcal{H}_{\frac{p^k-1}{2}}$ and  $|\mathcal{C}|=p^{mn(2\lambda-1) p^k}$}.

\vskip 2mm\par
  (iii-3-2) \textit{$(i_1,s_1)=(\lceil \frac{t}{2}\rceil,0)$, $(i_2,s_2)=(0,1)$, $(i_3,s_3)=(t,0)$,
$h(x)\in \mathcal{H}_{\lfloor \frac{t}{2}\rfloor}$ and  $|\mathcal{C}|=p^{mn(2\lambda p^k-t)}$, where $3\leq t\leq p^k-1$}.

\vskip 2mm\par
  (iv-1) \textit{$\lambda p^k-2$ cases}:
\textit{$(i_2,s_2)=(l_0+1,2l_1)$, $(i_3,s_3)=(l_0,2l_1+1)$,
$h(x)=0$ and  $|\mathcal{C}|=p^{mn(2(\lambda-l_1) p^k-2l_0-1)}$, where the pair $(l_0,l_1)$ of integers is given by one
of the following two cases}

\par
  $\diamond$ \textit{$(l_0,l_1)\neq (0,0)$, $0\leq l_0\leq p^k-1$ and $0\leq l_1\leq \lambda-2$}.

\par
  $\diamond$ \textit{$0\leq l_0\leq p^k-2$ and $l_1=\lambda-1$}.

\vskip 2mm\par
  (iv-2) \textit{$(\lambda p^k-3)p^{mn}$ cases}:
  \textit{$(i_1,s_1)=(l_0+1,2l_1)$, $(i_2,s_2)=(l_0,2l_1+1)$, $(i_3,s_3)=(l_0+2,2l_1)$,
$h(x)\in\mathcal{H}_1$ and  $|\mathcal{C}|=p^{2mn((\lambda-l_1) p^k-l_0-1)}$, where
the pair $(l_0,l_1)$ of integers is given by one
of the following two cases}

\par
  $\diamond$ \textit{$(l_0,l_1)\neq (0,0)$, $0\leq l_0\leq p^k-1$ and $0\leq l_1\leq \lambda-2$}.

\par
  $\diamond$ \textit{$0\leq l_0\leq p^k-3$ and $l_1=\lambda-1$}.

\vskip 2mm\par
  (iv-3) \textit{$\sum_{t=3}^{p^k}(\lambda p^k-1-t)p^{mn\lfloor\frac{t}{2}\rfloor}$ cases}:

\vskip 2mm\par
  (iv-3-1) \textit{$(i_1,s_1)=(l_0-\frac{p^k-1}{2},2(l_1+1))$, $(i_2,s_2)=(l_0,2l_1+1)$, $(i_3,s_3)=(l_0,2(l_1+1))$,
$h(x)\in\mathcal{H}_{\frac{p^k-1}{2}}$ and  $|\mathcal{C}|=p^{mn((2\lambda-2l_1-1) p^k-2l_0)}$, where
$\frac{p^k-1}{2}\leq l_0\leq p^k-1$
and $0\leq l_1\leq \lambda-2$}.

\vskip 2mm\par
  (iv-3-2) \textit{$(i_1,s_1)=(\frac{p^k+1}{2}+l_0,2l_1)$, $(i_2,s_2)=(l_0,2l_1+1)$, $(i_3,s_3)=(l_0,2(l_1+1))$,
$h(x)\in\mathcal{H}_{\frac{p^k-1}{2}}$ and  $|\mathcal{C}|=p^{mn((2\lambda-2l_1-1) p^k-2l_0)}$, where
$(l_0,l_1)\neq (0,0)$,
$0\leq l_0\leq \frac{p^k-3}{2}$ and $0\leq l_1\leq \lambda-2$}.

\vskip 2mm\par
  (iv-3-3) \textit{$(i_1,s_1)=(\lceil\frac{t}{2}\rceil+l_0,2l_1)$, $(i_2,s_2)=(l_0,2l_1+1)$, $(i_3,s_3)=(t+l_0,2l_1)$,
$h(x)\in\mathcal{H}_{\lfloor\frac{t}{2}\rfloor}$ and  $|\mathcal{C}|=p^{mn(2(\lambda-l_1)p^k-2l_0-t)}$, where
$(l_0,l_1)\neq (0,0)$,
$0\leq l_0\leq p^k-1-t$, $0\leq l_1\leq \lambda-1$ and $3\leq t\leq p^k-1$}.

\vskip 2mm\par
  (iv-3-4) \textit{$(i_1,s_1)=(\lceil\frac{t}{2}\rceil+l_0,2l_1)$, $(i_2,s_2)=(l_0,2l_1+1)$, $(i_3,s_3)=(t+l_0-p^k,2l_1+2)$,
$h(x)\in\mathcal{H}_{\lfloor\frac{t}{2}\rfloor}$ and  $|\mathcal{C}|=p^{mn(2(\lambda-l_1)p^k-2l_0-t)}$, where
$p^k-t\leq l_0\leq p^k-1-\lceil\frac{t}{2}\rceil$, $0\leq l_1\leq \lambda-2$
 and $3\leq t\leq p^k-1$}.

\vskip 2mm\par
  (iv-3-5) \textit{$(i_1,s_1)=(\lceil\frac{t}{2}\rceil+l_0-p^k,2l_1+2)$, $(i_2,s_2)=(l_0,2l_1+1)$, $(i_3,s_3)=(t+l_0-p^k,2l_1+2)$,
$h(x)\in \mathcal{H}_{\lfloor\frac{t}{2}\rfloor}$ and  $|\mathcal{C}|=p^{mn(2(\lambda-l_1)p^k-2l_0-t)}$, where
$p^k-\lceil\frac{t}{2}\rceil\leq l_0\leq p^k-1$, $0\leq l_1\leq \lambda-2$
 and $3\leq t\leq p^k-1$}.

\vskip 3mm\noindent
  {\bf Remark} When $p=3$ and $k=1$, there is no integer $t$ satisfying
$3\leq t\leq p^k-1=2$. Hence the subcase (iii-3-2) is wanting in Case (III), and there are only two subcases (iv-3-1) and (iv-3-2) in
Case (IV) of Theorem 5.1. For the special case of $\lambda=2$, a complete
description for $(\delta+\alpha u^2)$-constacyclic codes and their
dual codes over $\mathbb{F}_{3^m}[u]/\langle u^4\rangle$ of length $3n$
had been given (cf. [6] Corollary 4.6), where
$x^n-\delta_0$ is irreducible in $\mathbb{F}_{3^m}[x]$ and $\delta=\delta_0^3$.

\vskip 3mm\noindent
  {\bf Proof.} We only need to prove the cases (i-3) and (iv-3). The other conclusions
follows from Theorem 3.6, Corollary 3.7, Equations (1) and (8) immediately.

\par
  (i-3) As $(\lambda-1)p^k\leq s\leq \lambda p^k-3=(\lambda-1)p^k+(p^k-3)$, $s$ can be uniquely
expressed as $s=l_0+(\lambda-1)p^k$, where $l_0$ is an integer satisfying $0\leq l_0\leq p^k-3$.
This implies
$$\lceil\frac{\lambda p^k-s}{2}\rceil+s=\lceil\frac{p^k-l_0}{2}\rceil+l_0+(\lambda-1)p^k$$
and $0\leq \lceil\frac{p^k-l_0}{2}\rceil+l_0<\frac{p^k-l_0}{2}+1+l_0=\frac{p^k-l_0+2+2l_0}{2}\leq p^k-\frac{1}{2}<p^k$.
Hence by Equation (8) it follows that $\Psi((x^n-\delta_0)^s)=\alpha^{\lambda-1}(x^n-\delta_0)^{l_0}u^{2\lambda-2}$
and
$$\Psi((x^n-\delta_0)^{\lceil\frac{\lambda p^k-s}{2}\rceil+s})
=\alpha^{\lambda-1}(x^n-\delta_0)^{\lceil\frac{p^k-l_0}{2}\rceil+l_0}u^{2\lambda-2}.$$
As $0\leq \lfloor\frac{\lambda p^k-s}{2}\rfloor
=\lfloor\frac{p^k-l_0}{2}\rfloor\leq \frac{p^k-1}{2}$, by Equation (1)
we have $\Psi(h(x))=h(x)$ for any $h(x)\in \mathcal{A}/\langle (x^n-\delta_0)^{\lfloor\frac{\lambda p^k-s}{2}\rfloor}\rangle
=\mathcal{H}_{\lfloor\frac{\lambda p^k-s}{2}\rfloor}$.
Finally, the conclusion follows from $\mathcal{C}=\langle \Psi((x^n-\delta_0)^{\lceil\frac{\lambda p^k-s}{2}\rceil+s}) \Psi(h(x))+ u \Psi((x^n-\delta_0)^{s})\rangle$.

\par
  (iv-3) As $1\leq s\leq \lambda p^k-t-1=(\lambda-1)p^k+p^k-t-1$ and $3\leq t\leq p^k$,
we have one of the following cases.

\par
   {\bf Case 1} $t=p^k$. In this case, $1\leq s\leq \lambda p^k-p^k-1=(\lambda-2)p^k+(p^k-1)$,
$\lceil\frac{t}{2}\rceil=\frac{p^k+1}{2}$and $\lfloor\frac{t}{2}\rfloor=\frac{p^k-1}{2}$. Hence
$s$ can be uniquely
expressed as $s=l_0+l_1p^k$, where the pair $(l_0,l_1)$ of integers satisfying
$$0\leq l_0\leq p^k-1, \ 0\leq l_1\leq \lambda-2 \
{\rm and} \ (l_0,l_1)\neq (0,0).$$
Then it follows that $s+\lceil \frac{t}{2}\rceil=s+\lceil \frac{p^k}{2}\rceil=\frac{p^k+1}{2}+l_0+l_1p^k$. Hence we have
one of the following two subcases:

\par
  (iv-3-1) When $\frac{p^k-1}{2}\leq l_0\leq p^k-1$, $s+\lceil \frac{p^k}{2}\rceil=(\frac{p^k+1}{2}+l_0-p^k)+(l_1+1)p^k
=(l_0-\frac{p^k-1}{2})+(l_1+1)p^k$ and
$0\leq l_0-\frac{p^k-1}{2}\leq \frac{p^k-1}{2}$.
In this subcase, by Equations (1) and (8) we have
\begin{eqnarray*}
\mathcal{C}&=&\Psi(\langle  (x^n-\delta_0)^{s+p^k}, (x^n-\delta_0)^{s+\lceil \frac{p^k}{2}\rceil}h(x)+u(x^n-\delta_0)^s\rangle)\\
  &=&\langle  \alpha h(x)\cdot (x^n-\delta_0)^{l_0-\frac{p^k-1}{2}}u^{2(l_1+1)}
  +(x^n-\delta_0)^{l_0}u^{2l_1+1}, \\
  &&(x^n-\delta_0)^{l_0}u^{2(l_1+1)}\rangle
 \end{eqnarray*}
 with $|\mathcal{C}|=p^{mn((2\lambda-2l_1-1) p^k-2l_0)}$, where
$\alpha h(x)\in \alpha\mathcal{A}/\langle (x^n-\delta_0)^{\frac{p^k-1}{2}}\rangle=\mathcal{H}_{\frac{p^k-1}{2}}$.

\par
  (iv-3-2) When $0\leq l_0\leq \frac{p^k-3}{2}$, $s+\lceil \frac{p^k}{2}\rceil=(\frac{p^k+1}{2}+l_0)+l_1p^k$ and
$\frac{p^k+1}{2}+l_0\leq p^k-1$. In this subcase, by Equations (1) and (8) we have
\begin{eqnarray*}
\mathcal{C}&=&\Psi(\langle  (x^n-\delta_0)^{s+p^k}, (x^n-\delta_0)^{s+\lceil \frac{p^k}{2}\rceil}h(x)+u(x^n-\delta_0)^s\rangle)\\
  &=&\langle h(x)\cdot (x^n-\delta_0)^{\frac{p^k+1}{2}+l_0}u^{2l_1}
  +(x^n-\delta_0)^{l_0}u^{2l_1+1}, (x^n-\delta_0)^{l_0}u^{2(l_1+1)}\rangle
 \end{eqnarray*}
 with $|\mathcal{C}|=p^{mn((2\lambda-2l_1-1) p^k-2l_0)}$, where
$h(x)\in \mathcal{A}/\langle (x^n-\delta_0)^{\frac{p^k-1}{2}}\rangle=\mathcal{H}_{\frac{p^k-1}{2}}$.

\par
   {\bf Case 2} When $3\leq t\leq p^k-1$, then $p=3$ and $k\geq 2$ or $p\geq 5$. In this case, $1\leq s\leq \lambda p^k-t-1=(\lambda-1)p^k+(p^k-1-t)$ where $0\leq p^k-1-t\leq p^k-4$. Hence
$s$ can be uniquely
expressed as $s=l_0+l_1p^k$, where the pair $(l_0,l_1)$ of integers satisfies
one of the following two conditions:

\par
  ($\dag$) $(l_0,l_1)\neq (0,0)$, $0\leq l_0\leq p^k-1$ and $0\leq l_1\leq \lambda-2$;

\par
  ($\ddag$) $0\leq l_0\leq p^k-1-t$ and $l_1=\lambda-1$.

\noindent
 Moreover, we have that
$s+t=(t+l_0)+l_1p^k$ and $0\leq t+l_0\leq p^k-1$ when $0\leq l_0\leq p^k-1-t$; and
$s+t=(t+l_0-p^k)+(l_1+1)p^k$ and $0\leq t+l_0-p^k\leq p^k-1$ when $p^k-t\leq l_0\leq p^k-1$.
  Similarly, we have the following:

\par
 $\diamond$ $s+\lceil\frac{t}{2}\rceil=(\lceil\frac{t}{2}\rceil+l_0)+l_1p^k$ and $0\leq \lceil\frac{t}{2}\rceil+l_0\leq p^k-1$ when $0\leq l_0\leq p^k-1-\lceil\frac{t}{2}\rceil$;

\par
  $\diamond$ $s+\lceil\frac{t}{2}\rceil=(\lceil\frac{t}{2}\rceil+l_0-p^k)+(l_1+1)p^k$ and $0\leq \lceil\frac{t}{2}\rceil+l_0-p^k\leq p^k-1$ when $p^k-\lceil\frac{t}{2}\rceil\leq l_0\leq p^k-1$.

\par
  As stated above, we have one of the following three subcases:

\par
  (iv-3-3) $0\leq l_0\leq p^k-1-t$, $0\leq l_1\leq \lambda-1$ and $(l_0,l_1)\neq (0,0)$.
In this case, we have
\begin{eqnarray*}
\mathcal{C}&=&\Psi(\langle  (x^n-\delta_0)^{s+t}, (x^n-\delta_0)^{s+\lceil \frac{t}{2}\rceil}h(x)+u(x^n-\delta_0)^s\rangle)\\
  &=&\langle h(x)\cdot (x^n-\delta_0)^{l_0+\lceil\frac{t}{2}\rceil}u^{2l_1}
  +(x^n-\delta_0)^{l_0}u^{2l_1+1}, (x^n-\delta_0)^{l_0+t}u^{2l_1}\rangle
 \end{eqnarray*}
 with $|\mathcal{C}|=p^{mn(2(\lambda-l_1)p^k-2l_0-t)}$, where
$h(x)\in \mathcal{A}/\langle (x^n-\delta_0)^{\lfloor\frac{t}{2}\rfloor}\rangle=\mathcal{H}_{\lfloor\frac{ t}{2}\rfloor}$.

\par
  (iv-3-4) $p^k-t\leq l_0\leq p^k-1-\lceil\frac{t}{2}\rceil$ and $0\leq l_1\leq \lambda-2$.
In this case, we have
\begin{eqnarray*}
\mathcal{C}&=&\Psi(\langle  (x^n-\delta_0)^{s+t}, (x^n-\delta_0)^{s+\lceil \frac{t}{2}\rceil}h(x)+u(x^n-\delta_0)^s\rangle)\\
  &=&\langle h(x)\cdot (x^n-\delta_0)^{l_0+\lceil\frac{t}{2}\rceil}u^{2l_1}
  +(x^n-\delta_0)^{l_0}u^{2l_1+1}, (x^n-\delta_0)^{l_0+t-p^k}u^{2l_1+2}\rangle
 \end{eqnarray*}
 with $|\mathcal{C}|=p^{mn(2(\lambda-l_1)p^k-2l_0-t)}$, where
$h(x)\in \mathcal{H}_{\lfloor\frac{ t}{2}\rfloor}$.

\par
  (iv-3-5) $p^k-\lceil\frac{t}{2}\rceil\leq l_0\leq p^k-1$ and $0\leq l_1\leq \lambda-2$.
In this case, we have
\begin{eqnarray*}
\mathcal{C}&=&\Psi(\langle  (x^n-\delta_0)^{s+t}, (x^n-\delta_0)^{s+\lceil \frac{t}{2}\rceil}h(x)+u(x^n-\delta_0)^s\rangle)\\
  &=&\langle  \alpha h(x) \cdot(x^n-\delta_0)^{l_0+\lceil\frac{t}{2}\rceil-p^k}u^{2l_1+2}+(x^n-\delta_0)^{l_0}u^{2l_1+1},\\
  && (x^n-\delta_0)^{l_0+t-p^k}u^{2l_1+2}\rangle
 \end{eqnarray*}
 with $|\mathcal{C}|=p^{mn(2(\lambda-l_1)p^k-2l_0-t)}$, where
$\alpha h(x)\in \alpha\mathcal{H}_{\lfloor\frac{ t}{2}\rfloor}=\mathcal{H}_{\lfloor\frac{ t}{2}\rfloor}$.
\hfill $\Box$

\vskip 3mm
 \par
  Let $n=1$ and $\delta=\delta_0^{p^k}$. Then $x-\delta_0$ is irreducible in $\mathbb{F}_{p^m}[x]$
for any $\delta_0\in \mathbb{F}_{p^m}^\times$.
In the following, we give the explicit expressions for all distinct $(\delta+\alpha u^2)$-constacyclic code over $R$ of length $p^k$ over $R$ and their dual codes, where
$R=\mathbb{F}_{p^m}[u]/\langle u^{2\lambda}\rangle$ and $\lambda\geq 2$.
In this special case,
$\mathcal{T}_1=\mathcal{T}=\mathbb{F}_{p^m}$.

\par
   Let ${\cal C}$ be a $(\delta+\alpha u^2)$-constacyclic code over $R$ of length $p^k$.
Then ${\cal C}^{\bot}$ is a $(\delta+\alpha u^2)^{-1}$-constacyclic code of length $p^k$ over $R$, i.e. ${\cal C}^{\bot}$ is an ideal of the ring $R[x]/\langle x^{p^k}-(\delta+\alpha u^2)^{-1}\rangle$.
From $(\delta+\alpha u^2)^{-1}
=\delta^{-1}+\sum_{j=1}^{\lambda-1}(-1)^j\delta^{-(j+1)}\alpha^ju^{2j}$, by
$\delta_0^{p^k}=\delta$ and $x^{p^k}=(\delta+\alpha u^2)^{-1}$ we deduce that
\begin{equation}
(x-\delta_0^{-1})^{p^k}=\vartheta u^2, \ {\rm where} \
\vartheta=-\delta^{-2}\alpha+\sum_{j=2}^{\lambda-1}(-1)^j\delta^{-(j+1)}\alpha^ju^{2j-2}
\end{equation}
in $R[x]/\langle x^{p^k}-(\delta+\alpha u^2)^{-1}\rangle$ if $\lambda\geq 3$,
and $\vartheta=-\delta^{-2}\alpha$ if $\lambda=2$. Then $\vartheta$ is an
invertible element in $R$. Precisely, we have
$$\vartheta^{-1}=-\delta^2\alpha^{-1}-\delta u^2-\delta^2\alpha^{-1}(-\delta^{-1}\alpha)^{\lambda-1}u^{2\lambda-2}\in R.$$
In particular, $\vartheta^{-1}=-\delta^2\alpha^{-1}$ if $\lambda=2$. By $(\delta+\alpha u^2)x^{p^k}=1$ it follows that
\begin{equation}
x^{-l}=(\delta+\alpha u^2)x^{p^k-l} \ {\rm in} \ R[x]/\langle x^{p^k}-(\delta+\alpha u^2)^{-1}\rangle,
\ 0\leq l\leq p^k.
\end{equation}
 For any integer $l$, $1\leq l\leq p^k-1$, We will adopt the following notations
\begin{eqnarray*}
\mathcal{H}_l&=&\{\sum_{i=0}^{l-1}a_i(x-\delta_0)^i\mid a_0,a_1,\ldots,a_{l-1}\in \mathbb{F}_{p^m}\}\\
 &=&\{\sum_{i=0}^{l-1}h_ix^i\mid h_0,h_1,\ldots,h_{l-1}\in \mathbb{F}_{p^m}\}.
\end{eqnarray*}
In particular, we have $|\mathcal{H}_l|=p^{ml}$. For any $h(x)=\sum_{i=0}^{l-1}h_ix^i\in \mathcal{H}_l$, denote
$$\widehat{h}(x)=\tau(h(x))=h(x^{-1})=h_0+(\delta+\alpha u^2)\sum_{i=1}^{l-1}h_ix^{p^k-i}$$
in $R[x]/\langle x^{p^k}-(\delta+\alpha u^2)^{-1}\rangle$.
  Then from Theorems 5.1 and 4.2, by Equations (9) and (10) we deduce the following conclusion.

\vskip 3mm
\noindent
  {\bf Theorem 5.2} \textit{Using the notations above, denote
$$\pi=x-\delta_0, \ \widehat{\pi}=x-\delta_0^{-1} \ {\rm and} \
\varrho=-\delta_0(\delta+\alpha u^2)x^{p^k-1}.$$
Then all distinct
$(\delta+\alpha u^2)$-constacyclic codes over $R$ of length $p^k$ and their dual codes are given by
the following  four cases}.

\vskip 2mm\par
  (I) \textit{$1+p^{m}+\sum_{t=3}^{p^k}p^{m\lfloor\frac{t}{2}\rfloor}$ 1-generator codes}:

\vskip 2mm\par
  (i-1) \textit{$\mathcal{C}=\langle \pi^{p^k-1} u^{2\lambda-1}\rangle$ with $|\mathcal{C}|=p^{m}$},
\textit{and $\mathcal{C}^{\bot}=\langle u,\widehat{\pi}\rangle$}.

\vskip 2mm\par
  (i-2) \textit{$\mathcal{C}=\langle b\pi^{p^k-1} u^{2\lambda-2}+\pi^{p^k-2}u^{2\lambda-1}\rangle$  with $|\mathcal{C}|=p^{2m}$,
where $b\in \mathbb{F}_{p^m}$}, \textit{and $\mathcal{C}^{\bot}=\langle -b\varrho\widehat{\pi}+u,\widehat{\pi}^2\rangle$}.

\vskip 2mm\par
  (i-3)  \textit{$\mathcal{C}=\langle h(x)\cdot \pi^{\lceil\frac{p^k-l_0}{2}\rceil+l_0}u^{2\lambda-2}+ \pi^{l_0}u^{2\lambda-1}\rangle$ with $|\mathcal{C}|=p^{m(p^k-l_0)}$,
where $h(x)\in \mathcal{H}_{\lfloor\frac{p^k-l_0}{2}\rfloor}$
and $0\leq l_0\leq p^k-3$},
\textit{and}
$$\mathcal{C}^{\bot}=\langle  -\widehat{h}(x)\varrho^{\lceil\frac{p^k-l_0}{2}\rceil}\cdot \widehat{\pi}^{\lceil\frac{p^k-l_0}{2}\rceil}+u, \widehat{\pi}^{p^k-l_0}\rangle.$$

\par
  (II) \textit{$\lambda p^k+1$ 1-generator codes}:

\par
    $\diamond$ \textit{$\mathcal{C}=\langle 0\rangle$, in this case we have $\mathcal{C}^{\bot}=\langle 1\rangle$};

\par
 $\diamond$  \textit{$\mathcal{C}=\langle \pi^{l_0}u^{2l_1}\rangle$ with $|\mathcal{C}|=p^{2m((\lambda-l_1) p^k-l_0)}$, where $0\leq l_0\leq p^k-1$ and
$0\leq l_1\leq \lambda-1$}, \textit{and
$\mathcal{C}^{\bot}=\langle \widehat{\pi}^{p^k-l_0}u^{2(\lambda-l_1-1)}\rangle$}.

\vskip 2mm\par
  (III) \textit{$1+p^{m}+\sum_{t=3}^{p^k}p^{m\lfloor\frac{t}{2}\rfloor}$ 2-generator codes}:

\vskip 2mm\par
  (iii-1) \textit{$\mathcal{C}=\langle u,\pi\rangle$ with $|\mathcal{C}|=p^{mn(2\lambda p^k-1)}$},
\textit{and $\mathcal{C}^{\bot}=\langle \widehat{\pi}^{p^k-1} u^{2\lambda-1}\rangle$}.

\vskip 2mm\par
  (iii-2) \textit{$\mathcal{C}=\langle b\pi+u,\pi^2\rangle$ with $|\mathcal{C}|=p^{2mn(\lambda p^k-1)}$,
where $b\in \mathbb{F}_{p^m}$}, \textit{and $\mathcal{C}^{\bot}=\langle -b\varrho \cdot
\widehat{\pi}^{ p^k-1}u^{2\lambda-2}+\widehat{\pi}^{p^k-2}u^{2\lambda-1}\rangle$}.

\vskip 2mm\par
  (iii-3-1) \textit{$\mathcal{C}=\langle h(x)\cdot \pi^{\frac{p^k+1}{2}}+u,u^2\rangle$ with $|\mathcal{C}|=p^{m(2\lambda-1) p^k}$
where  $h(x)\in \mathcal{H}_{\frac{p^k-1}{2}}$},
\textit{and $\mathcal{C}^{\bot}=\langle -\widehat{h}(x)\varrho^{\frac{p^k+1}{2}}\cdot \widehat{\pi}^{\frac{p^k+1}{2}}u^{2\lambda-2}+ u^{2\lambda-1}\rangle$}.

\vskip 2mm\par
  (iii-3-2) \textit{$\mathcal{C}=\langle h(x)\cdot \pi^{\lceil \frac{t}{2}\rceil}+u,\pi^t\rangle$ with $|\mathcal{C}|=p^{m(2\lambda p^k-t)}$, where $3\leq t\leq p^k-1$ and
$h(x)\in \mathcal{H}_{\lfloor \frac{t}{2}\rfloor}$}, \textit{and}
$\mathcal{C}^{\bot}=\langle -\widehat{h}(x)\varrho^{\lceil\frac{t}{2}\rceil}\cdot
\widehat{\pi}^{p^k-t+\lceil\frac{t}{2}\rceil}u^{2\lambda-2}
+\widehat{\pi}^{p^k-t}u^{2\lambda-1}\rangle.$

\vskip 2mm\par
  (IV) \textit{$\lambda p^k-2+(\lambda p^k-3)p^{m}+\sum_{t=3}^{p^k}(\lambda p^k-1-t)p^{m\lfloor\frac{t}{2}\rfloor}$ codes}.

\vskip 2mm\par
  (iv-1) \textit{$\lambda p^k-2$ 2-generator codes}:

\par
  \textit{$\mathcal{C}=\langle \pi^{l_0+1}u^{2l_1}, \pi^{l_0}u^{2l_1+1}\rangle$ with $|\mathcal{C}|=p^{m(2(\lambda-l_1) p^k-2l_0-1)}$, where the pair $(l_0,l_1)$ of integers is given by one
of the following two cases}

\par
  $\diamond$ \textit{$(l_0,l_1)\neq (0,0)$, $0\leq l_0\leq p^k-1$ and $0\leq l_1\leq \lambda-2$};

\par
  $\diamond$ \textit{$0\leq l_0\leq p^k-2$ and $l_1=\lambda-1$},

\noindent
  \textit{and $\mathcal{C}^{\bot}=\langle  \widehat{\pi}^{p^k-l_0}u^{2(\lambda-l_1-1)}, \widehat{\pi}^{ p^k-l_0-1}u^{2(\lambda-l_1-1)+1}\rangle$}.

\vskip 2mm\par
  (iv-2) \textit{$(\lambda p^k-3)p^{m}$ 2-generator codes}:

\par
  \textit{$\mathcal{C}=\langle  b\pi^{l_0+1}u^{2l_1}+\pi^{l_0}u^{2l_1+1}, \pi^{l_0+2}u^{2l_1}\rangle$ with
  $|\mathcal{C}|=p^{2m((\lambda-l_1) p^k-l_0-1)}$,
where $b\in \mathbb{F}_{p^m}$ and the pair $(l_0,l_1)$ of integers is given by one
of the following two cases}

\par
  $\diamond$ \textit{$(l_0,l_1)\neq (0,0)$, $0\leq l_0\leq p^k-1$ and $0\leq l_1\leq \lambda-2$};

\par
  $\diamond$ \textit{$0\leq l_0\leq p^k-3$ and $l_1=\lambda-1$},

\noindent
  \textit{and}
$\mathcal{C}^{\bot}=\langle
-b\varrho\cdot \widehat{\pi}^{p^k-l_0-1}u^{2(\lambda-l_1-1)}+\widehat{\pi}^{p^k-l_0-2}u^{2(\lambda-l_1-1)+1},
\widehat{\pi}^{p^k-l_0}u^{2(\lambda-l_1-1)}\rangle.$

\vskip 2mm\par
  (iv-3) \textit{$\sum_{t=3}^{p^k}(\lambda p^k-1-t)p^{m\lfloor\frac{t}{2}\rfloor}$ 2-generator codes}:

\vskip 2mm\par
  (iv-3-1) \textit{$\mathcal{C}=\langle  h(x)\cdot\pi^{l_0-\frac{p^k-1}{2}}u^{2(l_1+1)}
  +\pi^{l_0}u^{2l_1+1}, \pi^{l_0}u^{2(l_1+1)}\rangle$ with
$|\mathcal{C}|=p^{m((2\lambda-2l_1-1) p^k-2l_0)}$, where
$h(x)\in \mathcal{H}_{\frac{p^k-1}{2}}$, $\frac{p^k-1}{2}\leq l_0\leq p^k-1$
and $0\leq l_1\leq \lambda-2$}, \textit{and}
$\mathcal{C}^{\bot}=\langle \widehat{\pi}^{p^k-l_0}u^{2(\lambda-l_1-1)}, -\vartheta^{-1}\widehat{h}(x)\varrho^{\frac{p^k+1}{2}}\cdot \widehat{\pi}^{p^k+\frac{p^k+1}{2}-l_0}
u^{2(\lambda-l_1-2)}
+\widehat{\pi}^{p^k-l_0}u^{2(\lambda-l_1-2)+1}\rangle$.

\vskip 2mm\par
  (iv-3-2) \textit{$\mathcal{C}=\langle  h(x)\cdot \pi^{\frac{p^k+1}{2}+l_0}u^{2l_1}
  +\pi^{l_0}u^{2l_1+1}, \pi^{l_0}u^{2(l_1+1)}\rangle$ with the number of codewords
$|\mathcal{C}|=p^{m((2\lambda-2l_1-1) p^k-2l_0)}$, where
$h(x)\in \mathcal{H}_{\frac{p^k-1}{2}}$, $(l_0,l_1)\neq (0,0)$,
$0\leq l_0\leq \frac{p^k-3}{2}$ and $0\leq l_1\leq \lambda-2$},
\textit{and}
\begin{eqnarray*}
\mathcal{C}^{\bot} &=& \langle -\vartheta\widehat{h}(x)\varrho^{\frac{p^k+1}{2}}\cdot \widehat{\pi}^{\frac{p^k+1}{2}-l_0}
u^{2(\lambda-l_1-1)}+\widehat{\pi}^{p^k-l_0}u^{2(\lambda-l_1-2)+1}, \\
 &&\widehat{\pi}^{p^k-l_0}u^{2(\lambda-l_1-1)}\rangle.
\end{eqnarray*}

\par
  (iv-3-3) \textit{$\mathcal{C}=\langle  h(x)\cdot \pi^{l_0+\lceil\frac{t}{2}\rceil}u^{2l_1}
  +\pi^{l_0}u^{2l_1+1}, \pi^{l_0+t}u^{2l_1}\rangle$
 with the number of codewords $|\mathcal{C}|=p^{m(2(\lambda-l_1)p^k-2l_0-t)}$, where
$h(x)\in \mathcal{H}_{\lfloor\frac{t}{2}\rfloor}$,
$(l_0,l_1)\neq (0,0)$, $0\leq l_0\leq p^k-1-t$, $0\leq l_1\leq \lambda-1$ and $3\leq t\leq p^k-1$},
\textit{and}
\begin{eqnarray*}
\mathcal{C}^{\bot} &=& \langle -\widehat{h}(x)\varrho^{\lceil \frac{t}{2}\rceil}\cdot \widehat{\pi}^{p^k-l_0-t+\lceil \frac{t}{2}\rceil}u^{2(\lambda-l_1-1)}+\widehat{\pi}^{p^k-l_0-t}u^{2(\lambda-l_1-1)+1},\\
 &&
\widehat{\pi}^{p^k-l_0}u^{2(\lambda-l_1-1)}\rangle.
\end{eqnarray*}

\par
  (iv-3-4) \textit{$\mathcal{C}=\langle  h(x)\cdot \pi^{l_0+\lceil\frac{t}{2}\rceil}u^{2l_1}
  +\pi^{l_0}u^{2l_1+1}, \pi^{l_0+t-p^k}u^{2l_1+2}\rangle$
with the number of codewords $|\mathcal{C}|=p^{m(2(\lambda-l_1)p^k-2l_0-t)}$, where
$h(x)\in \mathcal{H}_{\lfloor\frac{t}{2}\rfloor}$,
$p^k-t\leq l_0\leq p^k-1-\lceil\frac{t}{2}\rceil$, $0\leq l_1\leq \lambda-2$
 and $3\leq t\leq p^k-1$},
\textit{and}
\begin{eqnarray*}
\mathcal{C}^{\bot} &=& \langle -\vartheta \widehat{h}(x)\varrho^{\lceil \frac{t}{2}\rceil}\cdot \widehat{\pi}^{p^k-l_0-t+\lceil \frac{t}{2}\rceil}u^{2(\lambda-l_1-1)}+\widehat{\pi}^{2p^k-l_0-t}u^{2(\lambda-l_1-2)+1},\\
 &&
\widehat{\pi}^{p^k-l_0}u^{2(\lambda-l_1-1)}\rangle.
\end{eqnarray*}

\par
  (iv-3-5) \textit{$\mathcal{C}=\langle  h(x)\cdot \pi^{l_0+\lceil\frac{t}{2}\rceil-p^k}u^{2l_1+2}
 +\pi^{l_0}u^{2l_1+1}, \pi^{l_0+t-p^k}u^{2l_1+2}\rangle$
 with $|\mathcal{C}|=p^{m(2(\lambda-l_1)p^k-2l_0-t)}$, where
$h(x)\in \mathcal{H}_{\lfloor\frac{t}{2}\rfloor}$,
$p^k-\lceil\frac{t}{2}\rceil\leq l_0\leq p^k-1$, $0\leq l_1\leq \lambda-2$
 and $3\leq t\leq p^k-1$}, \textit{and}
\begin{eqnarray*}
\mathcal{C}^{\bot} &=& \langle -\vartheta^{-1}\widehat{h}(x)\varrho^{\lceil \frac{t}{2}\rceil}\cdot \widehat{\pi}^{2p^k-l_0-t+\lceil \frac{t}{2}\rceil}u^{2(\lambda-l_1-2)}+\widehat{\pi}^{2p^k-l_0-t}u^{2(\lambda-l_1-2)+1},\\
 &&
\widehat{\pi}^{p^k-l_0}u^{2(\lambda-l_1-1)}\rangle.
\end{eqnarray*}

\par
  \textit{Therefore, the number of codes over $R$ of length $p^k$
is equal to}
$N_{(p^m,2\lambda,p^k,1)}$ $=\sum_{l=0}^{\frac{p^k-1}{2}}\left(1+2\lambda p^k-4l\right)p^{lm}.$

\vskip 3mm \noindent
  {\bf Proof.} Let $i$ be an integer, $p^k\leq i\leq \lambda p^k-1$. Then there exists
a unique pair $(t,l)$ of integers such that
$i=tp^k+l$ where $t\geq 1$ and $0\leq l\leq p^k-1$. From Equation (9) we deduce that
\begin{equation}
\widehat{\pi}^i=(x-\delta_0^{-1})^i=((x-\delta_0^{-1})^{p^k})^t(x-\delta_0^{-1})^l
=(\vartheta u^2)^t\widehat{\pi}^l=\vartheta^t\widehat{\pi}^lu^{2t},
\end{equation}
and $\tau(\pi)=\tau(x-\delta_0)=x^{-1}-\delta_0=(-\delta_0x^{-1})(x-\delta_0^{-1})=\varrho\widehat{\pi}$
 in the ring $R[x]/\langle x^{p^k}-(\delta+\alpha u^2)^{-1}\rangle$. Then the conclusions follow from
Theorems 5.1, 4.2, 3.6 and Equation (11).
\hfill $\Box$

\vskip 3mm \noindent
  {\bf Remark} Using Equations (9)--(11), a more explicit expression for the dual code $\mathcal{C}^{\bot}$ can be
given. Here, we omit these works in order to save space.

\vskip 3mm\par
  Finally, we list all $(2+3u^2)$-constacyclic codes of length $5$
over the ring $R=\mathbb{F}_5[u]/\langle u^4\rangle$
and their dual codes by Theorem 5.2.

\par
   As $p=5$, $m=k=1$ and $\lambda=2$, the number of $(2+3u^2)$-constacyclic codes of length $5$ over $\mathbb{F}_5[u]/\langle u^4\rangle$ is equal to
$N_{(5,4,5,1)}=21+17\cdot 5+13\cdot 5^2=431$.

\vskip 2mm\noindent
  Moreover, we have $\alpha=3$ and $\delta=2$. It is clear that $\delta_0=2$
satisfying $\delta_0^5=\delta$, $\delta_0^{-1}=3$, $\vartheta=3$, $-\vartheta^{-1}=3$, $(2+3u^2)^{-1}=3+3u^2$ and

\vskip 2mm\par
 $\diamond$ $\mathcal{H}_2=\{h_0+h_1x\mid h_0,h_1\in \mathbb{F}_5\}$ with $|\mathcal{H}_2|=5^2$,
and $\mathcal{H}_1=\mathbb{F}_5$.

\vskip 2mm\par
 $\diamond$ $x^{-l}=(2+3u^2)x^{5-l}$ for any $1\leq l\leq 4$, and $-2x^{-1}=(1+4u^2)x^4$.

\vskip 2mm\par
 $\diamond$ $h(x^{-1})=h_0+(2+3u^2)h_1x^4$ for any $h(x)=h_0+h_1x\in \mathcal{H}_2$.

\vskip 2mm\par
  In order to save space, we adopt the following notations:
$\mathcal{R}=R[x]/\langle x^5-(2+3 u^2)\rangle$,
$\widehat{\mathcal{R}}=R[x]/\langle x^5-(2+3 u^2)^{-1}\rangle$ and

\vskip 2mm\par
  $\pi=x-2$, $\widehat{\pi}=x-3$, $\varrho=(1+4u^2)x^4$, $h=h(x)\in \mathcal{H}_2$ and $\widehat{h}=h(x^{-1})$.

\vskip 2mm\noindent
  Then $\pi^{5}=3u^2$ and $x^{5}=2+3u^2$ in $\mathcal{R}$;
$\widehat{\pi}^{5}=3u^2$, $x^{5}=3+3u^2$,
$$\tau(\pi)=x^{-1}-2=(-2x^{-1})(x-3)=\varrho \widehat{\pi} \
{\rm and} \ (1+u^2)\varrho^5=1 \ {\rm in} \ \widehat{\mathcal{R}}.$$
By Theorem 5.2, all $431$ $(2+3u^2)$-constacyclic codes $\mathcal{C}$  over $R$ of length $5$
and their dual codes $\mathcal{C}^{\bot}$ are given by the following table, where $\mathcal{C}^{\bot}$
is a $(3+3u^2)$-constacyclic code  over $R$ of length $5$.
{\footnotesize
\begin{center}
\begin{tabular}{llll|l}\hline
 N     &  Case &   $\mathcal{C}$  & $|\mathcal{C}|$  & $\mathcal{C}^{\bot}$  \\ \hline
$1$   & i-1 & $\langle \pi^4u^3\rangle$ & $5$  & $\langle u,\pi\rangle$ \\
$5$   & i-2 & $\langle b\pi^4u^2+\pi^3u^3\rangle$ ($b\in \mathbb{F}_{5}$)& $5^2$ & $\langle-b\varrho\pi+u,\pi^2\rangle$ \\
$25$  & i-3-1 & $\langle h\pi^{3}u^{2}+ u^{3}\rangle$
      & $5^{5}$
      & $\langle -\widehat{h}\varrho^{3}\widehat{\pi}^{3}+u, u^2\rangle$ \\
$25$ & i-3-2 & $\langle h\pi^{3}u^{2}+ \pi u^{3}\rangle$
     & $5^{4}$
     & $\langle -\widehat{h}\varrho^{2}\widehat{\pi}^{2}+u,
\widehat{\pi}^{4}\rangle$ \\
$5$  & i-3-3 & $\langle b\pi^{4}u^{2}+ \pi^{2}u^{3}\rangle$ ($b\in \mathbb{F}_{5}$)
     & $5^{3}$
     & $\langle -b\varrho^{2}\widehat{\pi}^{2}+u,
\widehat{\pi}^{3}\rangle$ \\
\hline
 $1$ & ii-1 & $\langle 0\rangle$ & $1$ & $\langle 1\rangle$ \\
 $10$ & ii-2 & $\langle \pi^{l_0}u^{2l_1}\rangle$ & $5^{20-10l_1-2l_0}$ & $\langle \widehat{\pi}^{5-l_0}u^{2-2l_1}\rangle$ \\
      &      & $0\leq l_0\leq 4$, $l_1\in\{0,1\}$  &  &  \\
\hline
$1$  & iii-1 & $\langle u,\pi\rangle$ &  $5^{19}$  & $\langle \widehat{\pi}^{4}u^3\rangle$ \\
$5$   & iii-2 & $\langle b\pi+u,\pi^2\rangle$ ($b\in \mathbb{F}_{5}$)
      &  $5^{18}$  & $\langle -b\varrho\widehat{\pi}^{4}u^2+\widehat{\pi}^{3}u^3\rangle$ \\
$25$  & iii-3-1 & $\langle h\pi^{3}+u,u^2\rangle$   & $5^{15}$
      & $\langle -\widehat{h}\varrho^{3}\widehat{\pi}^{3}u^{2}+ u^{3}\rangle$ \\
$25$  & iii-3-2 & $\langle h\pi^{2}+u,\pi^4\rangle$  & $5^{16}$
      & $\langle -\widehat{h}\varrho^{2}
\widehat{\pi}^{3}u^{2}
+ \widehat{\pi}u^{3}\rangle$ \\
$5$   & iii-3-3 & $\langle b\pi^{2}+u,\pi^3\rangle$ ($b\in\mathbb{F}_5$) & $5^{17}$
      & $\langle -b\varrho^{2}
\widehat{\pi}^{4}u^{2}
+ \widehat{\pi}^{2}u^{3}\rangle$ \\
\hline
$3$ & iv-1-1 & $\langle \pi^{l_0+1}, \pi^{l_0}u\rangle$ ($1\leq l_0\leq 3$) & $5^{20-2l_0-1}$
     & $\langle\widehat{\pi}^{5-l_0}u^{2}, \widehat{\pi}^{4-l_0}u^{3}\rangle$ \\
$1$  & iv-1-2 & $\langle u^2, \pi^{4}u\rangle$ & $5^{11}$
     & $\langle\widehat{\pi}u^{2}, u^{3}\rangle$ \\
$1$  & iv-1-3 & $\langle \pi u^{2}, u^{3}\rangle$ & $5^{9}$
     & $\langle u^2, \widehat{\pi}^{4}u\rangle$ \\
$3$  & iv-1-4 & $\langle \pi^{5-l}u^{2}, \pi^{4-l}u^{3}\rangle$ ($1\leq l\leq 3$) & $5^{2l+1}$
     & $\langle \widehat{\pi}^{l+1}, \widehat{\pi}^{l}u\rangle$  \\
\hline
$5$ & iv-2-1 & $\langle  bu^2+\pi^{4}u, \pi u^2\rangle$ ($b\in\mathbb{F}_5$) & $5^{10}$
     & $\langle  -bu^{2}+\varrho^4\widehat{\pi}^{4}u, \widehat{\pi}u^{2}\rangle$ \\
$15$ & iv-2-2 & $\langle  b\pi^{l_0+1}+\pi^{l_0}u, \pi^{l_0+2}\rangle$ & $5^{18-2l_0}$
     & $\langle-b\varrho\widehat{\pi}^{4-l_0}u^{2}+\widehat{\pi}^{3-l_0}u^{3},$ \\
     &        & $b\in\mathbb{F}_5$, $1\leq l_0\leq 3$ &  & $\widehat{\pi}^{5-l_0}u^{2}\rangle$\\
$15$ & iv-2-3 & $\langle  b\pi^{4-l}u^{2}+\pi^{3-l}u^{3}, \pi^{5-l}u^{2}\rangle$ & $5^{2l+2}$
     & $\langle -b\varrho \widehat{\pi}^{l+1}+\widehat{\pi}^{l}u, \widehat{\pi}^{l+2}\rangle$ \\
     &          & $b\in\mathbb{F}_5$, $1\leq l\leq 3$ & &  \\
\hline
$75$   & iv-3-1 & $\langle h\pi^{l_0-2}u^{2}+\pi^{l_0}u, \pi^{l_0}u^{2}\rangle$ & $5^{15-2l_0}$
       & $\langle  -\widehat{h}\varrho^{3}\widehat{\pi}^{3-l_0}u^2+\widehat{\pi}^{5-l_0}u,$ \\
       &      &    $l_0\in\{2,3\}$  &  & $\widehat{\pi}^{5-l_0}u^{2}\rangle$  \\
       &   & $\langle h\pi^{2}u^{2}+\pi^{4}u, \pi^{4}u^{2}\rangle$  & $5^{7}$
       & $\langle  3\widehat{h}\varrho^{3}\widehat{\pi}^{4}+\widehat{\pi}u, \widehat{\pi}u^{2}\rangle$ \\
$25$   & iv-3-2 &  $\langle  h\pi^{4}+\pi u, \pi u^{2}\rangle$ &  $5^{13}$
       & $\langle  2\widehat{h}\varrho^{3}\widehat{\pi}^{2}
         u^{2}+\widehat{\pi}^{4}u, \widehat{\pi}^{4}u^{2}\rangle$ \\
$40$   & iv-3-3 &  $\langle b\pi^{3}u^{2l_1}+\pi u^{2l_1+1}, \pi^{4}u^{2l_1}\rangle$  &  $5^{15-10l_1}$
       & $\langle -b\varrho^{2}\widehat{\pi}^{3}u^{2-2l_1}+\widehat{\pi}u^{3-2l_1}, $ \\
       &      & $b\in \mathbb{F}_5$, $l_1\in\{0,1\}$  &  & $\widehat{\pi}^{4}u^{2-2l_1}\rangle$\\
       &   &  $\langle  b\pi^{2}u^{2} +u^{3}, \pi^{3}u^{2}\rangle$ ($b\in \mathbb{F}_5$) &  $5^{7}$
       & $\langle  -b\varrho^{2}\widehat{\pi}^{4}+\widehat{\pi}^2u, u^2\rangle $ \\
       &    & $\langle  h\pi^{2}u^{2}+u^{3}, \pi^{4}u^{2}\rangle$  & $5^6$
       & $\langle  -\widehat{h}\varrho^{2} \widehat{\pi}^{3}+\widehat{\pi}u, u^{2}\rangle$ \\
$55$    & iv-3-4 &  $\langle  b\pi^{4}+\pi^{2}u, u^{2}\rangle$ ($b\in \mathbb{F}_5$) &  $5^{13}$
       & $\langle -b\varrho^{2}\widehat{\pi}^{2}u^{2}+u^3, \widehat{\pi}^{3}u^{2}\rangle$ \\
       &    &  $\langle  h\pi^{3}+\pi u, u^{2}\rangle$   & $5^{14}$
       & $\langle  -\widehat{h}\varrho^{2}\widehat{\pi}^2 u^{2}+u^3, \widehat{\pi}^{4}u^{2}\rangle$ \\
       &    &  $\langle  h\pi^{4}+\pi^2 u, \pi u^{2}\rangle$   & $5^{12}$
       & $\langle 2\widehat{h}\varrho^2\widehat{\pi} u^{2}+\widehat{\pi}^4u, \widehat{\pi}^{3}u^{2}\rangle$ \\
$60$   & iv-3-5  & $\langle b\pi^{l_0-3}u^{2}+\pi^{l_0}u, \pi^{l_0-2}u^{2} \rangle$ & $5^{17-2l_0}$
       &  $\langle -b\varrho^{2}\widehat{\pi}^{4-l_0}u^2+\widehat{\pi}^{7-l_0}u,$ \\
       &  & $b\in\mathbb{F}_5$, $l_0\in \{3,4\}$ &  &  $\widehat{\pi}^{5-l_0}u^{2}\rangle$ \\
       &   & $\langle h u^{2}
          +\pi^{3}u, \pi^{2}u^{2}\rangle$  & $5^{10}$
          & $\langle -\widehat{h}\varrho^{2}u^2+\widehat{\pi}^{3}u, \widehat{\pi}^{2}u^{2}\rangle$ \\
       &   & $\langle h \pi u^{2}+\pi^{4}u, \pi^{3}u^{2}\rangle$  & $5^{8}$
          & $\langle 3\widehat{h}\varrho^{2}\widehat{\pi}^{4}+\widehat{\pi}^{2}u, \widehat{\pi}u^{2}\rangle$ \\
\hline
\end{tabular}
\end{center}  }

\noindent
where $N$ is the number of codes in the same case
and $h=h(x)\in\mathcal{H}_2$.


\section{Conclusions and further research}\label{}
\noindent
We give an explicit representation and a complete description for all distinct $(\delta+\alpha u^2)$-constacyclic codes of
length $np^k$ over $R=\mathbb{F}_{p^m}[u]/\langle u^e\rangle$ and their dual codes, where $p$ is an odd prime number, $e$ is an even integer satisfying $e\geq 4$ and
${\rm gcd}(p,n)=1$. Our further interest
is to consider the minimum distance of each $(\delta+\alpha u^2)$-constacyclic code over $R$
of length $p^k$.

\par
  The proof of Lemma 3.5 in the paper depends on that $p$ is odd. Open
problems and further researches in this area include characterizing $(\delta+\alpha u^2)$-constacyclic
codes of length $2^kn$ over $\mathbb{F}_{2^m}[u]/\langle u^e\rangle$ for any odd positive integer $n$,
and integers $e,k$ satisfying $e\geq 4$ and $k\geq 2$ respectively.



\vskip 3mm \noindent {\bf Acknowledgments}
 Part of this work was done when Yonglin Cao was visiting Chern Institute of Mathematics, Nankai University, Tianjin, China. Yonglin Cao would like to thank the institution for the kind hospitality. This research is
supported in part by the National Natural Science Foundation of
China (Grant Nos. 11671235, 11801324, 61571243, 11471255), the Shandong Provincial Natural Science Foundation, China (Grant No. ZR2018BA007) and the Scientific
Research Fund of Hunan Provincial Key Laboratory of Mathematical Modeling and Analysis in Engineering(No. 2018MMAEZD04).
H.Q. Dinh and S. Sriboonchitta are grateful for the Center of Excellence in Econometrics, Chiang Mai University, Thailand, for partial financial support.





\end{document}